\documentclass[12pt,]{article}
\usepackage{lmodern}
\usepackage{amssymb,amsmath}
\usepackage{ifxetex,ifluatex}
\usepackage{fixltx2e} 
\ifnum 0\ifxetex 1\fi\ifluatex 1\fi=0 
  \usepackage[T1]{fontenc}
  \usepackage[utf8]{inputenc}
\else 
  \ifxetex
    \usepackage{mathspec}
  \else
    \usepackage{fontspec}
  \fi
  \defaultfontfeatures{Ligatures=TeX,Scale=MatchLowercase}
\fi
\IfFileExists{upquote.sty}{\usepackage{upquote}}{}
\IfFileExists{microtype.sty}{%
\usepackage{microtype}
\UseMicrotypeSet[protrusion]{basicmath} 
}{}
\usepackage[margin=1in]{geometry}
\usepackage{hyperref}
\hypersetup{unicode=true,
            pdftitle={Building and Testing Yield Curve Generators for P\&C Insurance},
            pdfauthor={Gary Venter -- Columbia University Department of Actuarial Sciences, and Kailan Shang -- Aon PathWise Solutions Group},
            pdfborder={0 0 0},
            breaklinks=true}
\usepackage{color}
\usepackage{fancyvrb}

\DefineVerbatimEnvironment{Highlighting}{Verbatim}{commandchars=\\\{\}}
\usepackage{framed}
\definecolor{shadecolor}{RGB}{248,248,248}
\newenvironment{Shaded}{\begin{snugshade}}{\end{snugshade}}

\newcommand{\CommentTok}[1]{\textcolor[rgb]{0.56,0.35,0.01}{\textit{#1}}}

\newcommand{\DataTypeTok}[1]{\textcolor[rgb]{0.13,0.29,0.53}{#1}}
\newcommand{\DecValTok}[1]{\textcolor[rgb]{0.00,0.00,0.81}{#1}}

\newcommand{\KeywordTok}[1]{\textcolor[rgb]{0.13,0.29,0.53}{\textbf{#1}}}
\newcommand{\NormalTok}[1]{#1}
\newcommand{\OperatorTok}[1]{\textcolor[rgb]{0.81,0.36,0.00}{\textbf{#1}}}
\newcommand{\OtherTok}[1]{\textcolor[rgb]{0.56,0.35,0.01}{#1}}

\newcommand{\StringTok}[1]{\textcolor[rgb]{0.31,0.60,0.02}{#1}}

\usepackage{graphicx,grffile}
\makeatletter
\def\maxwidth{\ifdim\Gin@nat@width>\linewidth\linewidth\else\Gin@nat@width\fi}
\def\maxheight{\ifdim\Gin@nat@height>\textheight\textheight\else\Gin@nat@height\fi}
\makeatother
\setkeys{Gin}{width=\maxwidth,height=\maxheight,keepaspectratio}
\IfFileExists{parskip.sty}{%
\usepackage{parskip}
}{
\setlength{\parindent}{0pt}
\setlength{\parskip}{6pt plus 2pt minus 1pt}
}
\ifx\paragraph\undefined\else
\let\oldparagraph\paragraph
\renewcommand{\paragraph}[1]{\oldparagraph{#1}\mbox{}}
\fi
\ifx\subparagraph\undefined\else
\let\oldsubparagraph\subparagraph
\renewcommand{\subparagraph}[1]{\oldsubparagraph{#1}\mbox{}}
\fi

\let\rmarkdownfootnote\footnote%
\def\footnote{\protect\rmarkdownfootnote}

\usepackage{titling}

  \title{Building and Testing Yield Curve Generators for P\&C Insurance}
    \pretitle{\vspace{\droptitle}\centering\huge}
  \posttitle{\par}
    \author{Gary Venter -- Columbia University Department of Actuarial Sciences, and
Kailan Shang -- Aon PathWise Solutions Group}
    \preauthor{\centering\large\emph}
  \postauthor{\par}
    \date{}
    \predate{}\postdate{}
  
\usepackage{setspace}
\begin{document}
\maketitle

\textbf{Abstract} Interest-rate risk is a key factor for
property-casualty insurer capital. P\&C companies tend to be highly
leveraged, with bond holdings much greater than capital. For GAAP
capital, bonds are marked to market but liabilities are not, so shifts
in the yield curve can have a significant impact on capital. Yield-curve
scenario generators are one approach to quantifying this risk. They
produce many future simulated evolutions of the yield curve, which can
be used to quantify the probabilities of bond-value changes that would
result from various maturity-mix strategies. Some of these generators
are provided as black-box models where the user gets only the projected
scenarios. One focus of this paper is to provide methods for testing
generated scenarios from such models by comparing to known
distributional properties of yield curves.

Typically regulators, security analysts, and customers focus on one to
three-year timeframes for capital risk. This is much different than
risk-management in other financial institutions, where the focus is on
how much markets can move from one day's close to the next day's
opening. Those institutions trade continuously when the markets are
open, and manage risk with derivatives. P\&C insurers, on the other
hand, hold bonds to maturity and manage cash-flow risk by matching asset
and liability flows. Derivative pricing and stochastic volatility are of
little concern over the relevant time frames. This requires different
models and model testing than what is common in the broader financial
markets.

To complicate things further, interest rates for the last decade have
not been following the patterns established in the sixty years following
WWII. We are now coming out of the period of very low rates, yet are
still not returning to what had been thought of as normal before that.
Modeling and model testing are in an evolving state while new patterns
emerge.

Our analysis starts with a review of the literature on interest-rate
model testing, with a P\&C focus, and an update of the tests for current
market behavior. We then discuss models, and use them to illustrate the
fitting and testing methods. The testing discussion does not require the
model-building section. We do try to make the modeling more accessible
to actuarial modelers, compared to our source papers in the financial
literature. Code for MCMC estimation is included at the CAS GitHub site.
Model estimation is getting easier as the software advances, and
interested actuaries, who often have a better feel for the application
needs than do financial modelers, can use this to fit their own
yield-curve generators.


\textbf{1 A historical look}\\
In this Section 1 we look at a long-term history of US interest rates
for perspective. Section 2 reviews the literature on properties of yield
curves for testing models, and updates the properties in the light of
recent data. Section 3 introduces affine models, which often meet most
of the yield-curve tests. In section 4 we fit some models. Section 5
then illustrates the tests by applying them to the fitted models.
Appendix 1 covers some more general affine models, and Appendix 2
addresses fitting models by Markov-chain Monte Carlo (MCMC).

\begin{figure} 
\centerline{\includegraphics[width=1.2\textwidth]{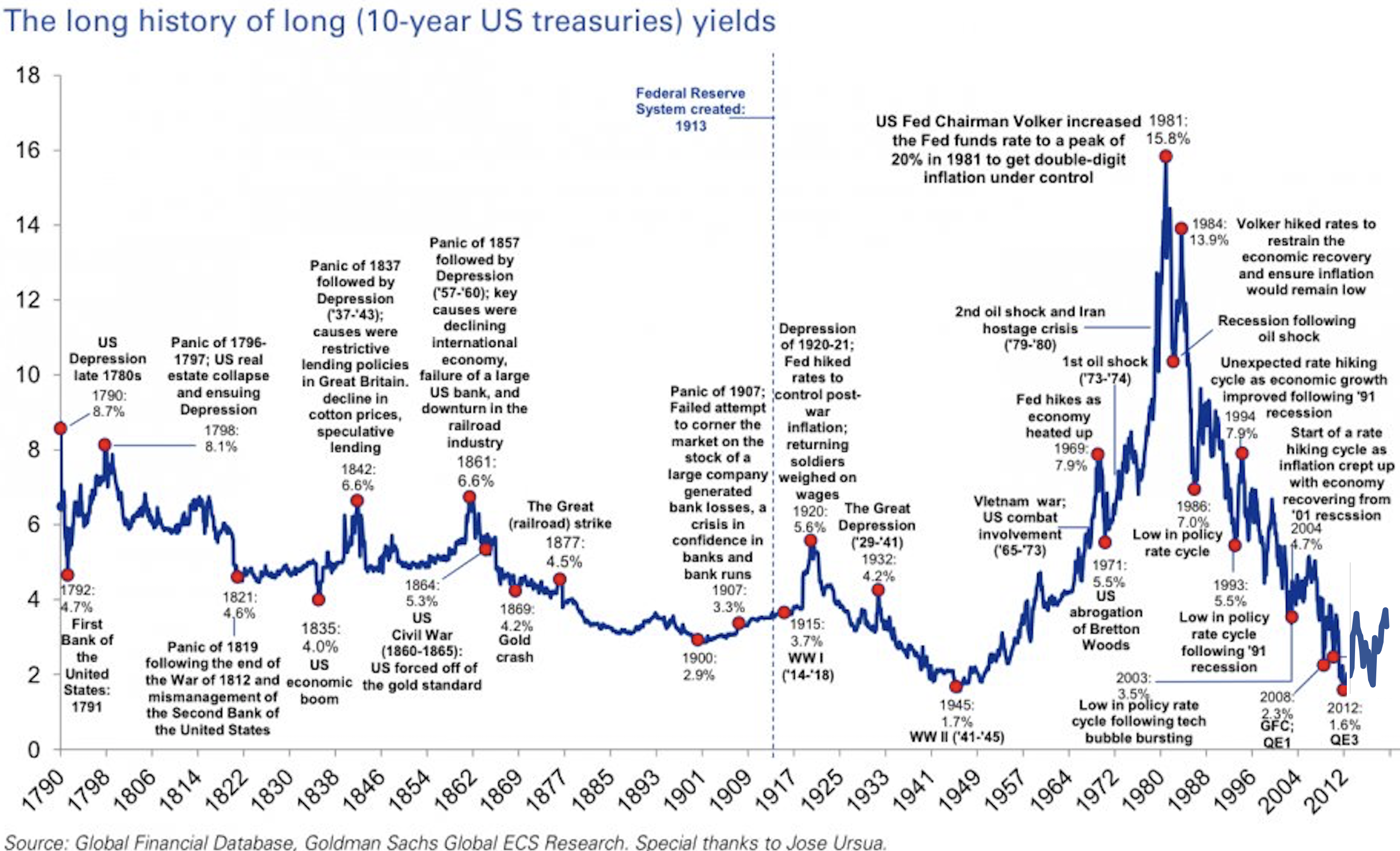}}
\caption{History of Ten Year US Bond Rates}
\end{figure}

Figure 1 graphs long Treasury bond yields for 1790--2018. It is hard to
identify a long-term equilibrium rate. Before the Civil War, the rate
gravitated around 6\%, but from 1870 to 1970, it was rarely above 4\% --
and it has returned to similar levels since about 2010. The postwar
period of higher rates -- say 1972 to 2007 -- was considered the new
normal at the time, and many economists who spent their careers in those
years still consider it normal.

The longer history calls that normality into question. Some economists
are now pointing out historical peculiarities of the postwar period.
Piketty (2014) reports a few such findings. For instance, that period
included a now-completed massive rebuilding of global assets, 50\% of
which, in monetary terms, were destroyed in the world wars. It was also
a time of steadily increasing productivity, since diminished. Both of
those elements boosted economic expansion, wages, and interest rates.
Absent the postwar period, the current situation looks more like a
continuation of the 1870--1970 levels than a return to 1990.

The graph also shows long periods of rising or falling rates. The rate
generally declined for 39 years starting in 1861, then rose for 20
years, and then declined for 25 years, to 1.7\% in 1945. Then it
increased for 36 years, with some fluctuations, and dropped again for 31
years, getting to 1.8\% in 2012.

\begin{figure}
\centering
\includegraphics{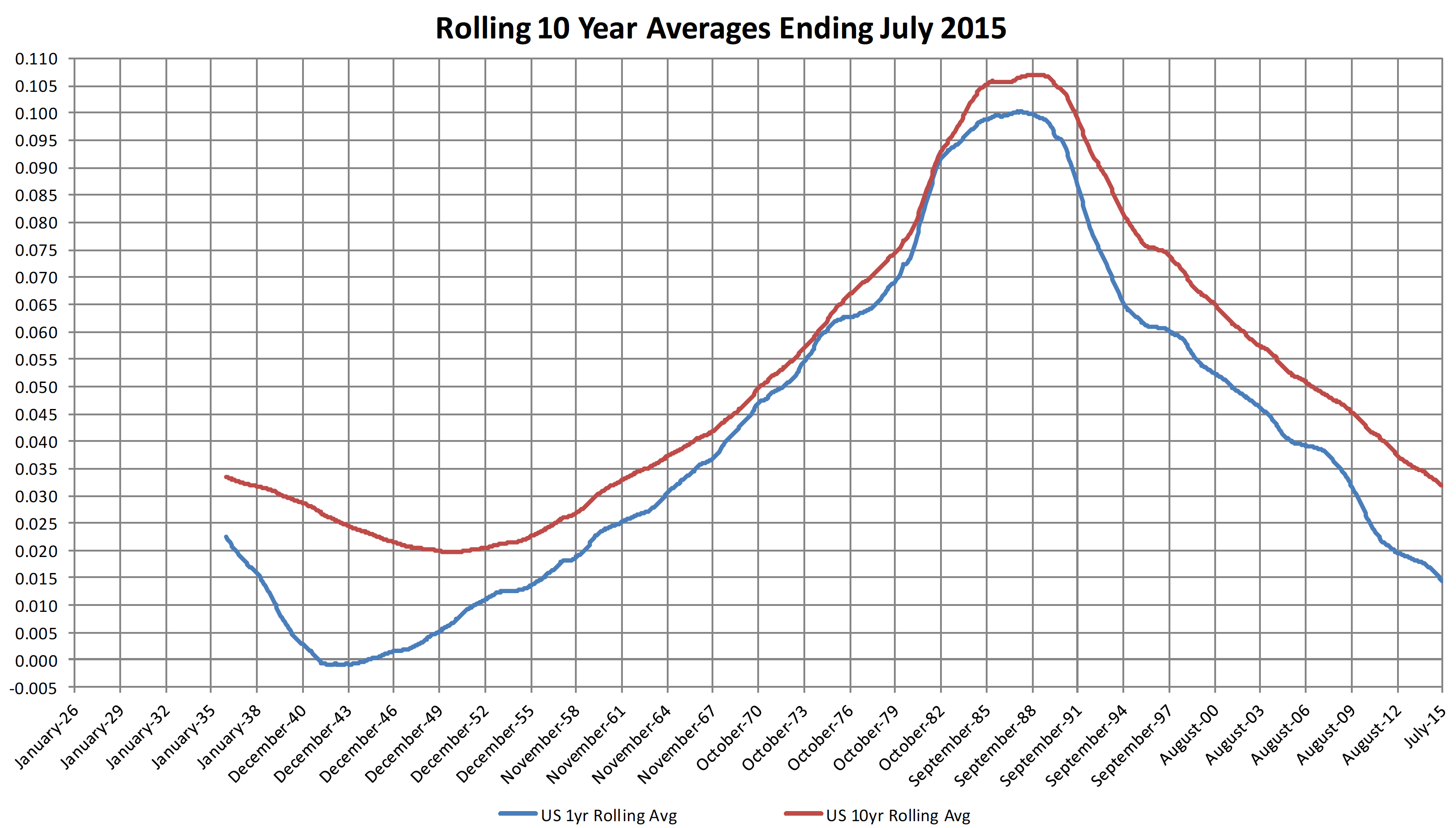}
\caption{Ten-Year Rolling Average 1 and 10 Year Rates\label{fig1}}
\end{figure}

Figures 2 and 3, from Pedersen et al. (2016), the SoA ESG report, show
rolling average one and ten year rates and their standard deviations
since 1936. The one-year rates are lower, but their standard deviations
are higher, compared to the ten-year rates. The standard deviations
loosely follow the rates. The Cox-Ingersoll-Ross (CIR) interest rate
model assumes that for the shortest rates, the standard deviation is
proportional to the rates. The gap between the one-year and ten-year
rates was high in the early 1940s, much like it was recently. In the
1940s the one-year rates were even lower than they were in the 2010s and
actually became negative for a few years.

\begin{figure}
\centering
\includegraphics{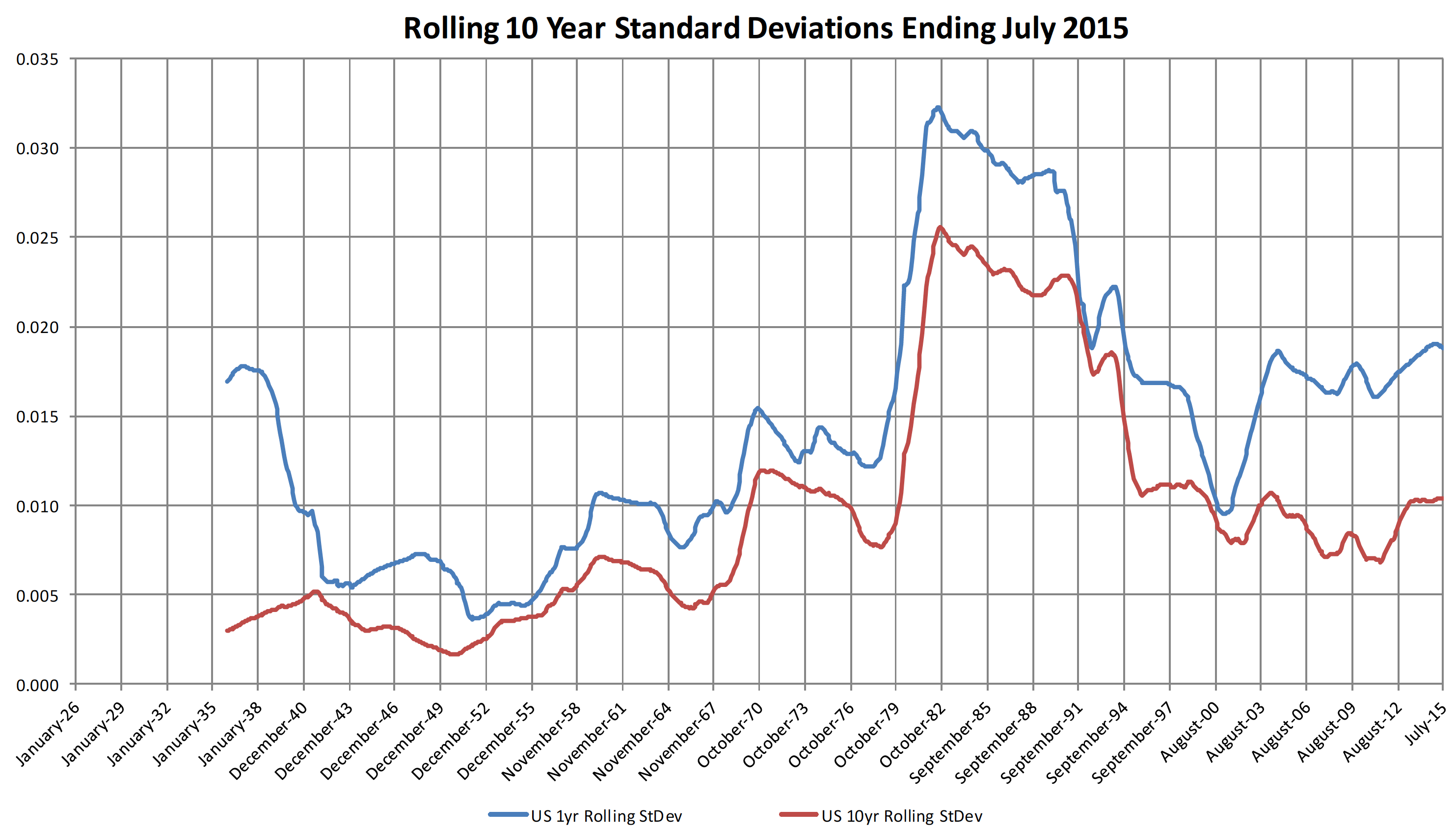}
\caption{Ten-Year Rolling Average 1 and 10 Year Standard
Deviations\label{fig1}}
\end{figure}

\textbf{2 Historical model tests and current updates}\\
Several tests have been proposed for yield-curve models. We discuss
proposed tests from three papers, and review these with recent data. The
period from 2008 to 2018 had unusually low rates -- comparable to those
from 1938 to 1955. With the recent changes in economic conditions, it
seems unlikely that such rates would return over the next several
decades. But the historically unusually high rates from 1970 -- 2000 had
specific causes that also seem unlikely to repeat. This makes it
difficult to select a relevant time period for measuring rate
properties. What we do below is look at how the properties have been
evolving over time in order to come up with reasonable criteria for
model behavior. Many of the tests would be performed on scenarios
simulated from the fitted models, but a few use fits to the data.

\textbf{\emph{2a Tests from the Feldh\H{u}tter paper on fitting affine
models}}\\
Feldhütter (2016) explores how well affine models fit historical
properties of interest rates.

\textbf{\emph{2a.1 Moments by maturity}}\\
Table 1 shows his exhibit of moments for US Treasury maturities of 1 --
5 years, using monthly observations from 1952:6 to 2004:12. This
includes the unusually high rates of the 1970s and 80s. What it shows is
mean rates that increase by maturity and volatilities that decrease.
Higher moments are positive but small and also decline a bit by
maturity.

\begin{table}[]
\centering
\caption{Moments of Treasury Yields}
\begin{tabular}{lrrrrr}
maturity        & 1 & 2 & 3 & 4 & 5\\
mean            & 5.60  & 5.81 & 5.98 & 6.11 & 6.19 \\
standard deviation  & 0.50 & 0.43 & 0.40 & 0.39 & 0.36 \\
skewness        & 0.83 & 0.79 & 0.78 & 0.77 & 0.77 \\
excess kurtosis & 0.77 & 0.57 & 0.51 & 0.44 & 0.35 \\
\end{tabular}
\end{table}

The modestly positive higher moments in Table 1 have been challenging
for models to match -- most give much higher or virtually zero higher
moments. If this holds up for more relevant data, it would be an
important feature to capture for risk analysis.

The yield curve is usually upward sloping, and there are good reasons
for that. Investors need higher yields to lock up their funds in longer
rates, and to take a many-year risk of bond values going down due to
rates increasing. The curve gets flat or inverted when the Federal
Reserve raises short rates above the level at which the market is
trading the longer obligations. This seems reasonably likely to happen
over the next several years, so some curve inversions would be a good
thing to see in a simulated scenario set.

\begin{figure}
\centering
\includegraphics{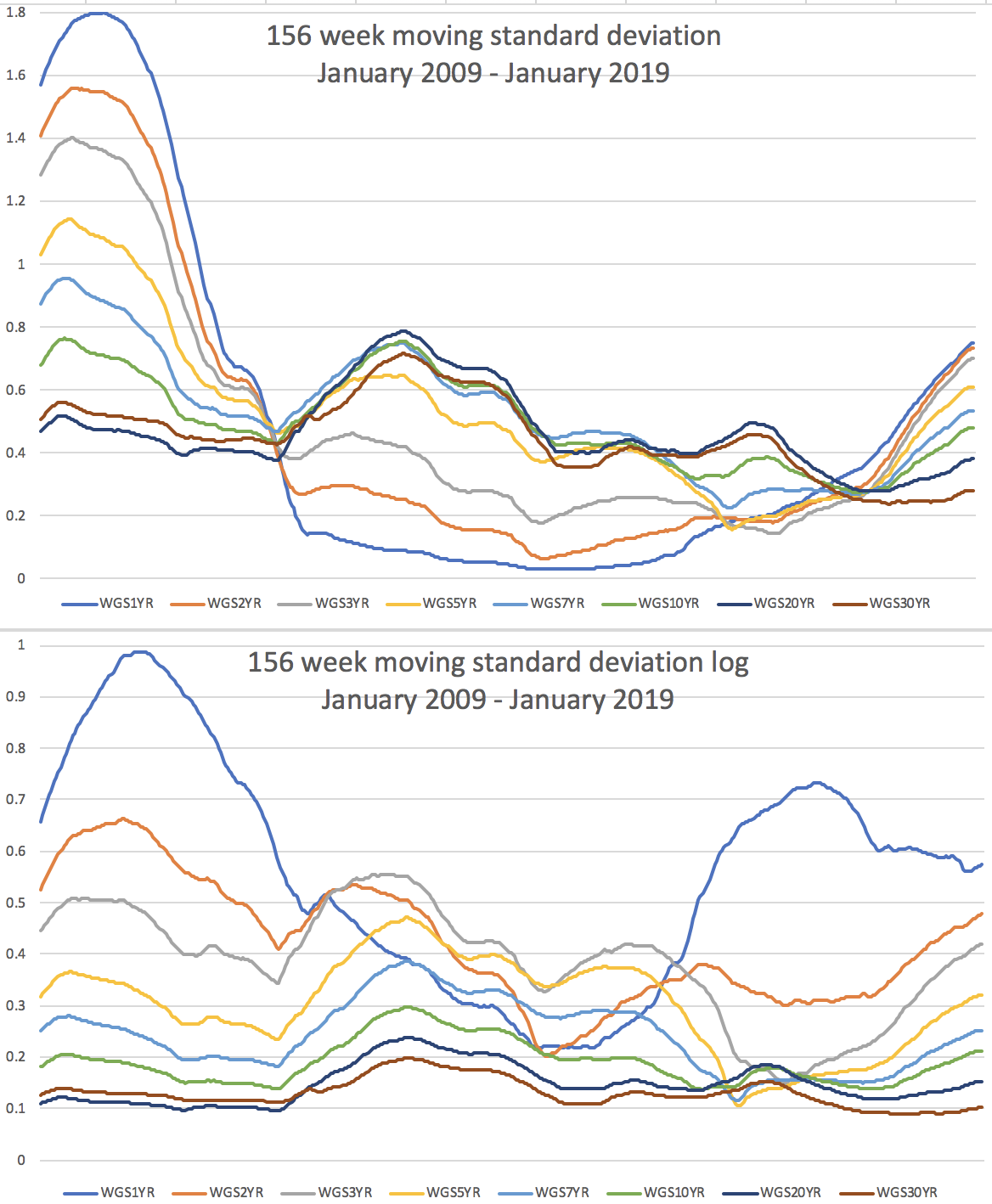}
\caption{Three-Year Moving Standard Deviation of Rates and Log Rates by
Maturity\label{fig1}}
\end{figure}

Although shorter rates are normally more volatile than are longer rates,
this pattern was reversed for six years beginning in August 2011. The
short rates were too low to have much absolute volatility. As an
alternative we looked at the volatility of the log of the rates.

Figure 4 shows 156-week moving standard deviations of rates and the log
of rates from 1/2009 to 1/2019. Before August 2011 and after August 2017
the usual pattern held, with volatility decreasing as maturities
increased. In between, the pattern was almost exactly reversed. The logs
of the rates show more consistent volatilities. Except for a short
period where rates longer than two years all had very little volatility,
the usual pattern is maintained for those maturities. The one and
two-year maturities always have higher log volatility than do the ten
and longer year maturities, and drop below the three-to-seven-year log
volatilities for only a relatively short period. Going forward it seems
that the modeled volatilities of the logs of the rates should decline
steadily with maturity. In the near term, it looks like the one-year log
volatilities have stabilized around 0.6, while the others show a bit of
an upward trend. Numbers a little higher than these would thus seem
reasonable targets for model testing.

Figure 5 shows the skewness of the five-year rate over fairly long
periods. It displays the rolling skewness of weekly rates over moving
500, 1000, and 1500 week periods, with the periods' ending points
starting in 1971. These correspond roughly to 10, 20, and 30 year
periods. The higher values in the earlier years appear to be due to the
very high rates in the early 1980s. There is no consistent skewness that
holds in general. There is a temporary spike due to the recovery from
very low rates (below 1\%) starting in 2016. These have since stabilized
around 2.5\%, so the spike seems to be over. For simulated rates going
forward, very little skewness would be a good result -- say below 0.25
but not too negative either.

\begin{figure}
\centering
\includegraphics{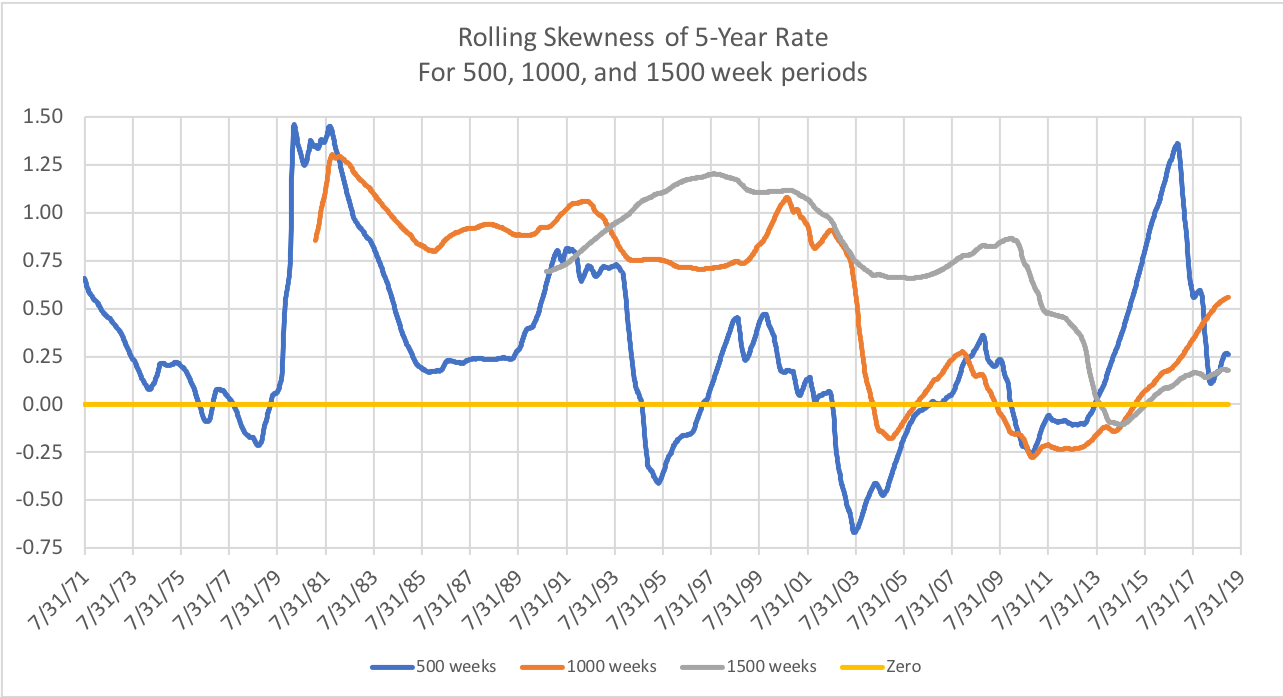}
\caption{500, 1000, and 100 Week Moving Skewness of Five-Year
Rates\label{fig1}}
\end{figure}

\textbf{\emph{2a.2 Volatility as related to levels of the rates}}\\
Feldhütter (2016) regresses the squared change in rates from month \(t\)
to \(t+1\) for each maturity against the level, slope, and curvature of
the yield curve at month \(t\). The level is the five-year rate, the
slope is the difference between the five and one-year rates, and the
curvature is the sum of the five and one-year rates less twice the
three-year rate, respectively. This measure of curvature is higher when
the midpoint of the curve is lower relative to the endpoints, so
quantifies upward curvature. Table 2 shows the coefficients for each
variable, along with their t-statistics, for each maturity.

The coefficients are small, as the squared change in rates is. The level
coefficients all have quite significant t-statistics, showing that the
volatility of rates is higher when the level of rates is. This effect is
greatest for the shorter maturities, since the volatilities are higher
for them. The other coefficients are not individually significant, but
the slope coefficients display a clear pattern of lower volatility for
the shorter maturities for steeper yield curves. Volatility is also
consistently but not significantly higher with less downward curvature.

\begin{table}[]
\centering
\caption{Volatility Regression Coefficients}
\begin{tabular}{lrrrrr}
maturity  & 1     & 2     & 3     & 4     & 5      \\
level     & 0.11  & 0.07  & 0.06  & 0.05  & 0.04    \\
t-statistic         & -5.4  & -5.4  & -5.5  & -6.9  & -7.6    \\
slope     & -0.14 & -0.08 & -0.03 & -0.02 & 0.02    \\
t-statistic          & 1.5   & 1.2   & 0.7   & 0.5   & -0.5   \\
curvature & 0.27  & 0.11  & 0.17  & 0.08  & 0.13    \\
t-statistic          & -1.4  & -0.8  & -1.7  & -1.0  & -2.1   \\

\end{tabular}
\end{table}

However, the previous results no longer hold up. Test found that these
coefficients were no longer significant, and sometimes had changed sign.
The very low rates did not act in accord with the earlier results. This
does not appear to be a useful test at this time, but should be reviewed
as new data emerges.

\textbf{\emph{2a.3 Annual rate changes by maturity related to curve
slope}}\\
The value of a bond would generally go up over time as its time to
maturity shortens. These changes have been found historically to be
stronger (bigger value increase, bigger yield drop) when the current
yield spread over the shortest yield is higher. Feldhütter (2016) does a
regression for this effect, following the methodology of Campbell and
Shiller (1991). With yield \(Y(t,n)\) at time \(t\) and time to maturity
\(n\), he finds the factors for a regression on the change in yield.
That is, he
estimates:\[Y(t+1,n-1) - Y(t,n) = const + factor_n \frac{Y(t,n)-Y(t,1)}{n-1} + res\]The
factors, shown in Table 3, are negative, which means that the drops in
yield are greater when the spread over the one-year yield is higher, and
are increasingly so as the maturity increases. This is a sort of mean
reversion that investors would be looking for. A longer bond that has a
particularly high spread relative to the shortest bond would be expected
to increase more in value during the next year. Campbell and Shiller
(1991) called this the excess-profits effect. Apparently the higher
spread led later to a greater decrease in the longer rate. This seems
counter-intuitive, as short rates are more volatile, so the wider spread
would be expected to narrow due to an increase in the short rate, not a
decrease in the long rate. Feldhütter (2016) also finds this effect
suspect, but confirms it empirically.

\begin{table}[]
\centering
\caption{Yield Change Regression}
\begin{tabular}{lrrrr}
maturity    & 2     & 3     & 4     & 5      \\
factor & -0.78 & -1.13 & -1.52 & -1.49   \\
t-statistic           & 1.4   & 1.8   & 2.2   & -2.0    \\
\end{tabular}
\end{table}

\begin{table}[]
\centering
\caption{Yield Change Slopes by Period}
\begin{tabular}{rrrrr}
Period & 4/53-12/78   & 1/84-12/08 & 1/09-1/19 & 4/53-1/19 \\
2Y     & -4.81 & -0.36      & -1.73     & -0.93     \\
3Y     & -4.63 & -0.34      & -2.54     & -1.18     \\
5Y     & 0.91   & 0.76       & -3.46     & -0.54     \\
7Y     & 1.18  & 0.65       & -5.02     & -0.66     \\
10Y    & 1.86  & 1.39       & -4.43     & -0.25    
\end{tabular}
\end{table}

Our Federal Reserve data does not have 4, 6, 8, or 9 year rates, so to
update the analysis we looked at the change over two years for bonds
that started as 5 or 7 years, and the change over three years for the 10
year bonds. Preliminary analysis suggested that the very high rates
around 1981 were distorting the results, but the very low rates of the
last ten years were also unique. For this reason, we did regressions
over 4 periods: the first 25 years in our data, from mid-1953 to 1978,
the 25 years from 1984 through 2008 (thus skipping 1978--1984), 2009 --
2019, and the entire 66 years. Table 4 shows the results.

The entire period shows all negative slopes, but generally getting less
negative by maturity. The two 25-year older periods show negative slopes
for the 2 and 3-year rates, with increasingly positive slopes for the
longer rates. Only the 10 recent years display the pattern found in
earlier studies.

The past decade has featured very low short rates, with very low
volatility. Thus in this period, a high spread over the 1-year rate
would indicate a higher long rate. Then for it to decrease is just a
form of mean reversion. The first two columns were 25-year periods of
either generally increasing or generally decreasing rates, but their
coefficients are similar. It could be that in those periods, higher
spreads occurred during expansionary times, where overall rates were
increasing more, or decreasing less, than when spreads were lower.

To test simulated rates going forward, negative coefficients would be
reasonable for 2 and 3 year rates. For longer rates, right now any
result could look plausible -- positive, negative, or not significant
coefficients. In the future, another look at the regressions would be
called for.

\textbf{\emph{2b Test from the Venter (2004) paper on ESGs for P\&C
companies}}

\textbf{\emph{2b.1 Yield spreads and the short rate}}\\
This paper presents a test of yield curve scenarios based on the way
longer yield spreads relate to the short rate. Since longer rates are
less volatile, they go up less when the short rate increases. Thus the
spread, say between the three and ten-year rates, would be expected to
decrease when the short rate increases, and widen again when it goes
back down. Figure 6 shows data on this from 1995 to 2011 with a
regression fit. The scatter around the regression line is fairly
consistent for three sub-periods shown.

\begin{figure}
\centering
\includegraphics{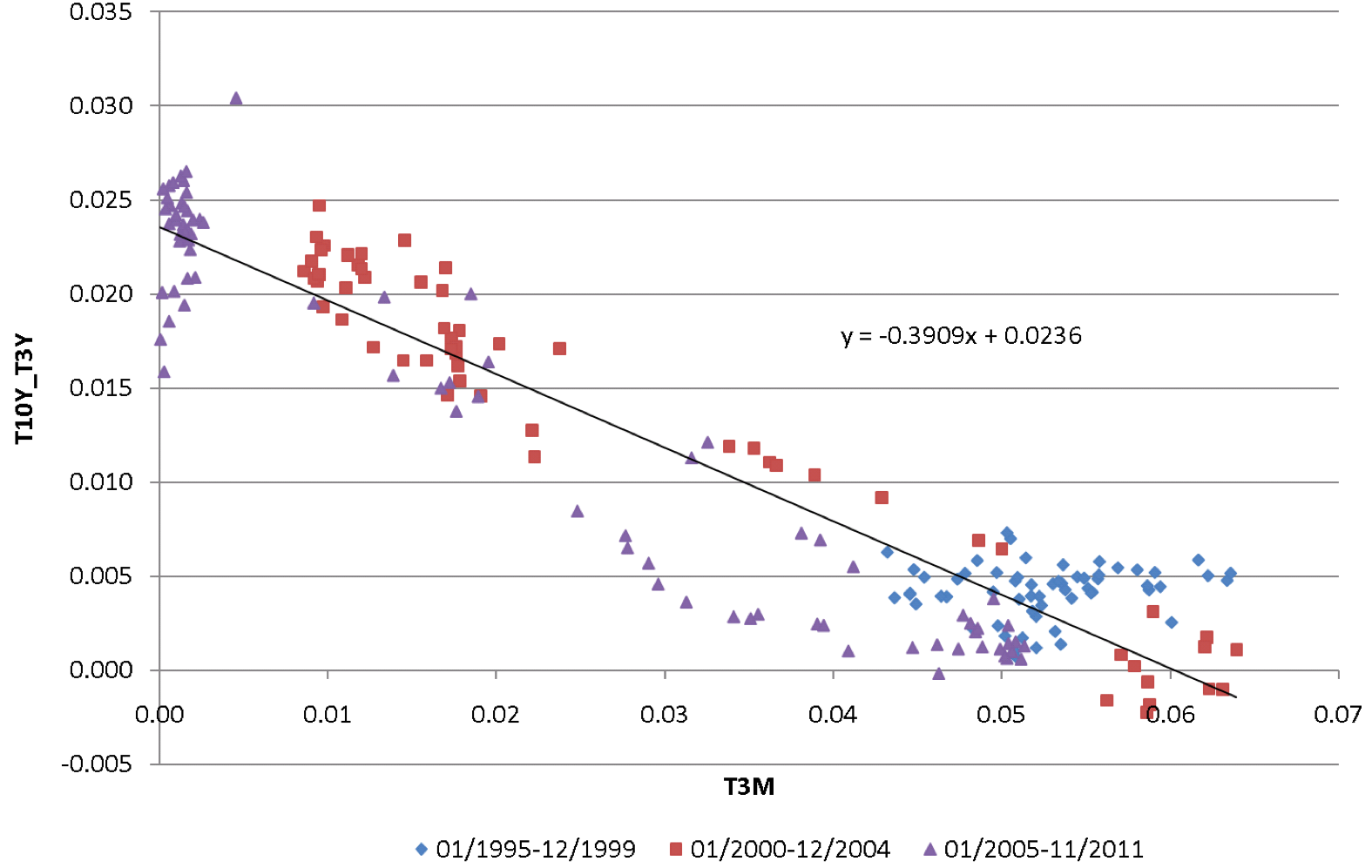}
\caption{Three-Year to Ten-Year Spreads as a Function of the Three-Month
Rate\label{fig1}}
\end{figure}

\begin{figure}
\centering
\includegraphics{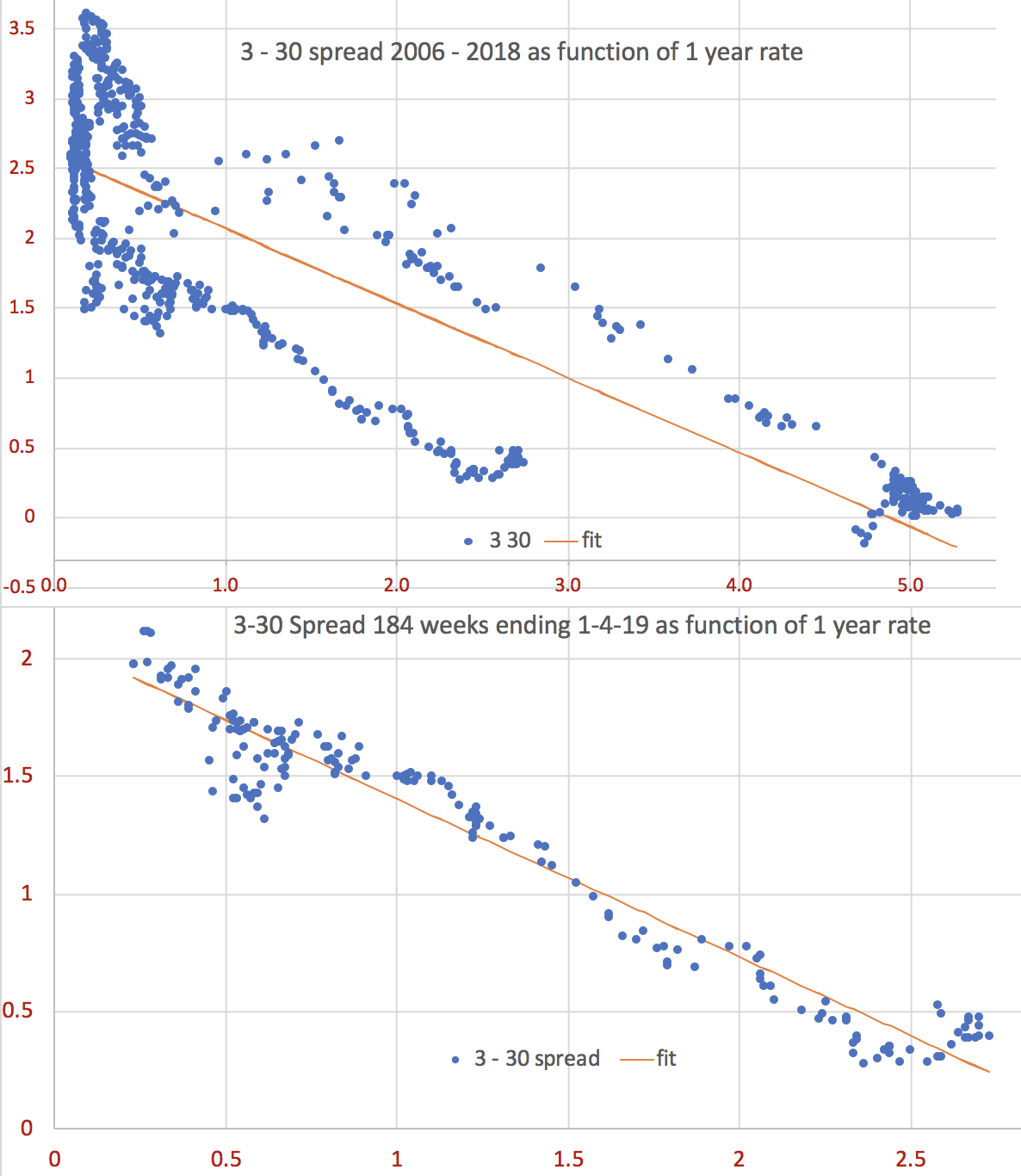}
\caption{3 -- 30 Year Spread as Function of 1-Year Rates\label{fig1}}
\end{figure}

Figure 7 graphs the 3-year -- 30-year spread as a function of the 1-year
rate for two periods -- 13 years starting in 2006, and about 3\(^1/_2\)
years ending 1/4/19, both with fitted lines. The longer range shows two
distinct periods, with similar slopes but different intercepts. The
second of these periods is what is fit in the bottom panel. The graphs
are similar to Figure 6, which shows the 3-10 spread for 1995-2001, so
the general long-term pattern has been continuing. The slope in the
upper panel is \(-0.53\), compared to \(-0.67\) in the lower panel. The
standard deviation around the slope for the longer period is 0.57, but
it is clearly less than this for the two sub-periods. It is 0.12 for the
later period.

To use this as a test for simulated scenarios, fitting the trend line
would be the starting point, then graphing the data and fitted line. An
eyeball test of the overall look of the graph compared to Figures 6 and
7 would be a check of the basic pattern. For the near future, a slope
similar to the recent period would be expected. For a longer projection,
any slope less than \(-0.5\) would seem reasonable. A flatter line would
indicate that the longer spreads do not compress much with increasing
shorter rates, which could arise from problems in the volatilities. For
a short-term projection, the spread around the line should be fairly
small -- below 0.2 perhaps -- but a wider spread would be reasonable for
a longer projection. Still, any spread above 0.6 would suggest that the
historical behavior is not reflected in the model.

\textbf{\emph{2c Tests from the Jagannathan, Kaplin, and Sun (2003)
paper about testing CIR models}}\\
This paper uses properties of yield curves for expanded model
goodness-of-fit tests on the sample data. It compares fits from
single-factor, two-factor, and three-factor CIR models. They try several
tests, but the two here show ways in which only the three-factor model
fits well. They have other tests involving option prices where none of
the CIR models work. But since bond option prices depend on stochastic
volatility, CIR would not be expected to work. This should not be a
problem for the time frames used in P\&C models, where the stochastic
volatilities average out.

\textbf{\emph{2c.1 Volatility by maturity}}\\
They compare volatility by maturity in the sample with that implied by
the fits. Figure 8 is their main result for that, and it shows a
reasonably good match for the three-factor model. A similar test for
simulated data is included in the Feldhütter (2016) discussion above.

\begin{figure}
\centering
\includegraphics{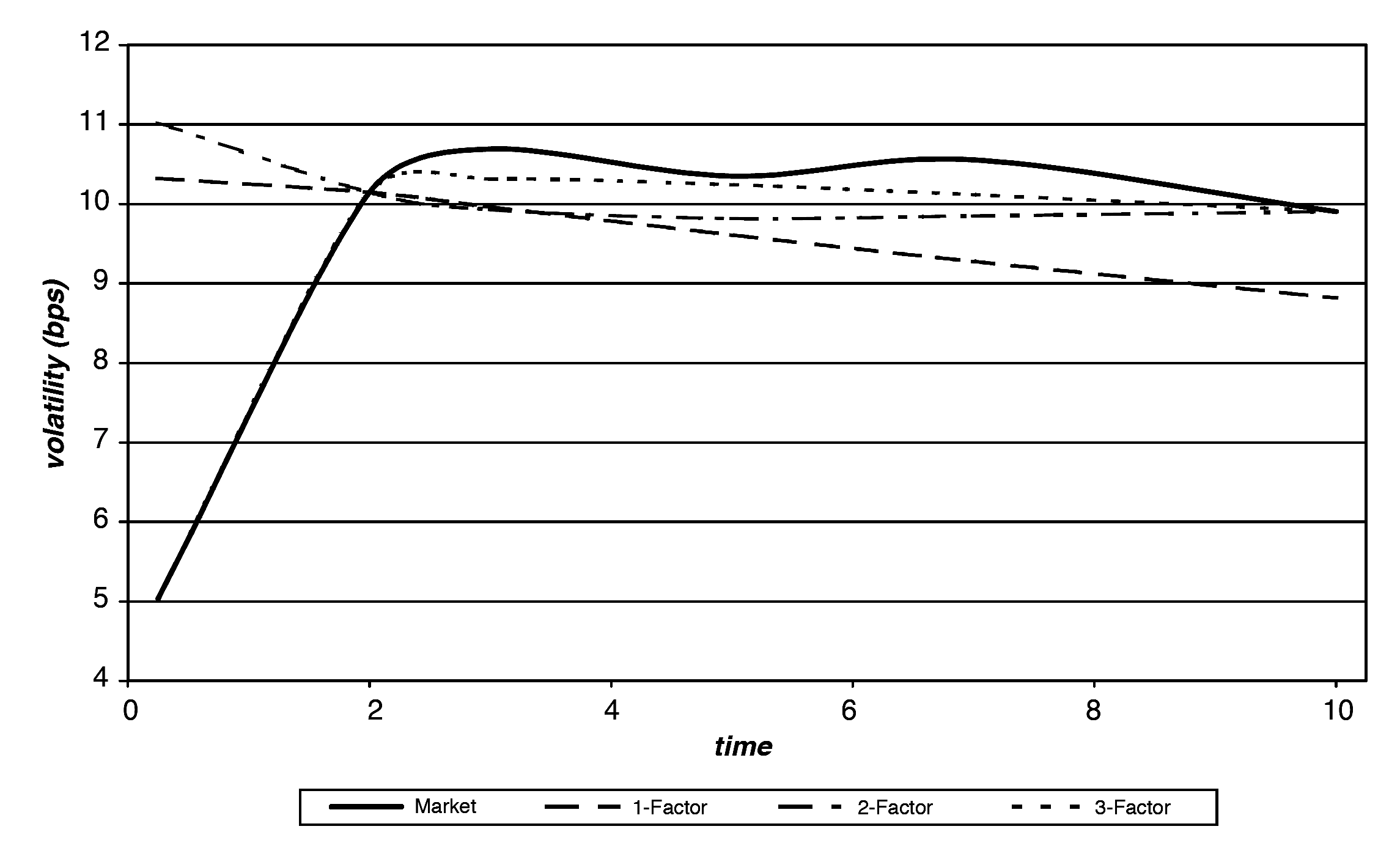}
\caption{Actual and Implied Volatilites by Maturity for each
Model\label{fig1}}
\end{figure}

\textbf{\emph{2c.2 Yield curve shapes as described by principal
component analysis (PCA)}}\\
Yield data comes in an array -- perhaps rows for each week of
observation and columns for each maturity. In their data there are six
maturities, so the yield curve at any date can be represented as a
vector in six-dimensional space. PCA rotates the axes of this space to
put most of the variation into a few dimensions. The first such
dimension -- i.e., the first principal component -- is defined by the
line between the two points that are furthest apart in this space. The
second axis is the longest line between two points that is perpendicular
to the first one, etc. Each axis gives a set of weights for the yields
at any date, where the sum product of the weights with the yields is the
coordinate on that axis. For interest rates, usually three principal
components -- often called level, slope, and curvature -- are enough to
explain most of the variation in the yield curves. That is, there is
very little variation in the remaining dimensions. There are standard
formulas that produce principal components and software to implement
them.

The weights for the first four components for their example are in Table
5, along with the total variance accounted for by each. The first three
explain over 99\% of the variation in yield curves. Then they look at
how well these three principal components can be approximated by the
fitted models, using a regression approach. They find that the
three-factor model provides a very good approximation to the three
principal components, while the other models do not. They conclude that
a three-factor CIR can capture the variety of yield curve shapes, but
two factors cannot.

With modern fitting, there would be fitted values for each observation,
so these could be used to compute fitted principal components for
comparison to the actual principal components. That would show how well
the fit is able to explain the observed variation in yield curves.

\begin{table}[]
\centering
\caption{Weights for the first four principal components of yield changes}
\begin{tabular}{lrrrrrrr}
Change in: & 3M     & 2Y     & 3Y     & 5Y     & 7Y    & 10Y     & \% Explained \\
Level      & 0.123  & 0.430  & 0.462  & 0.450  & 0.454 & 0.421   & 93.7         \\
Slope      & -0.866 & -0.273 & -0.112 & 0.081  & 0.223 & 0.326   & 3.6          \\
Curvature  & 0.482  & -0.601 & -0.288 & -0.035 & 0.355 & 0.4433  & 2.0          \\
Factor 4   & -0.010 & -0.606 & 0.662  & 0.235  & 0.016 & -0.3738 & 0.3         \\
\end{tabular}
\end{table}

The principal components are essentially two or three new yield curves
fit using the observed collection of yield curves. In the example from
Jagannathan, Kaplin, and Sun (2003), 94\% of the differences among
yields across the data set is explained by the first component, and over
97\% of the shape differences can be accounted for by regressing each
yield curve on the first two components. They stop with three
components, which explain 99.3\% of the variability.

For a recent historical dataset, we tried PCA on the actual data, and on
two fitted models, which were two and three-factor affine models. These
model the yields as linear combinations of partial short rates. Then for
each observed time point, we calculated the PCA-fitted yield curves from
the three sets of components. Figure 9 graphs the actual and fitted
curves at one particular time. This sample has a difficult set of yield
curves, due to a reversal in rates from 1 to 5 years. As can be seen,
the actual and historical PCA curves are practically identical, and the
three-factor model gives a reasonably good approximation. The two-factor
model has less flexibility and does not match this particular curve
shape.

\begin{figure}
\centering
\includegraphics{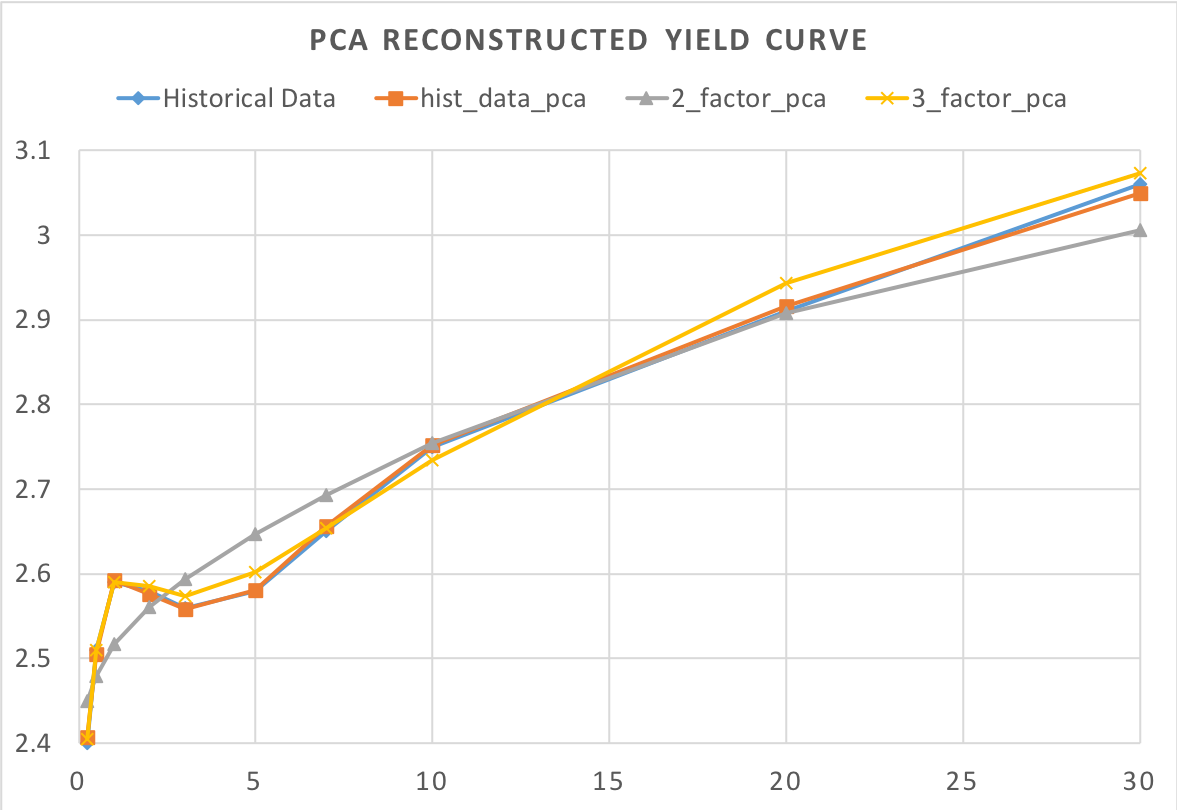}
\caption{Yield Curve from PCA \label{fig1}}
\end{figure}

What we found in doing this exercise, and should have known in advance,
is that the two-factor model fits only have two principal components,
and the three factor model's have three. That's because the fitted
values are linear combinations of two or three sets of short rates. Thus
the results of Jagannathan, Kaplin, and Sun (2003) -- that the
two-factor CIR does not fit the three principal components of actual
rates -- was more or less pre-ordained. As long as the third actual
component is non-trivial, a two-factor model will not have enough
variety of curve shapes.

This gives another way to do a realism test of simulated yield curves:
do a PCA on the set of curves and see if the third component explains
enough of the variance -- maybe 0.005 or more. An expanded test would be
to do a PCA on what is taken to be a relevant historical period, and use
that as a guideline on how significant the third component should be.

Another useful indicator of how well the model fits by maturity would be
to just compute an \(R^2\) statistic for each maturity. For each
maturity, that would be the total variance of the rates at the maturity
minus the variance of the model residuals, as a percentage of that total
variance. Table 6 illustrates this for the two example models. The
problems of the two-factor model show up in some specific maturities.

\begin{table}[]
\centering
\caption{R-Squared by Maturity}
\begin{tabular}{crrrrrrrrrr}
$R^2$                & 3M     & 6M     & 1Y     & 2Y     & 3Y     & 5Y     & 7Y     & 10Y    & 20Y    & 30Y    \\
Two-factor fit   & 95.7\% & 97.7\% & 77.6\% & 94.9\% & 95.6\% & 91.6\% & 96.5\% & 99.1\% & 91.7\% & 81.5\% \\
Three-factor fit & 99.4\% & 99.6\% & 97.9\% & 98.1\% & 99.2\% & 97.3\% & 99.6\% & 99.4\% & 98.6\% & 99.2\%
\end{tabular}
\end{table}

\textbf{3 Interest rate models}\\
Generally the academic literature focuses on arbitrage-free models. Some
actuarial models do not require that, depending on the applications
intended. If the risk of different investment strategies is to be
analyzed, having arbitrage possible could easily distort the
conclusions. If realistic yield curves are desired, the models should
also be arbitrage-free. But if rates for only one or two maturities are
to be projected long-term, such constraints might not be necessary. Here
we will stick to arbitrage-free models of the whole yield curve.

Lognormal models, like the Libor Market Model, produce more skewed
distributions of rates than the historical data supports. They are
popular for options' pricing, where the greater skewness may help match
market risk pricing of short-term volatility. That is not the main focus
of P\&C models, though, and so we will use models with normal residuals.

Most popular of these are the affine models. They build up the yield
curves from the short rates. In a single-factor model with short-rate
\(r(t)\) at time \(t\), every maturity \(\tau\) has a constant term
\(C(\tau)\) and a factor \(D(\tau)\) that do not change over time, with
fitted rate \(R(t,\tau) = C(\tau) + D(\tau) r(t)\). Thus changes in the
yield curve over time depend only on the changes in the short rate
\(r(t)\). This will make all the yield curves parallel, which is
unrealistic.

Multi-factor model postulate short-rate components \(r_j(t)\) for each
factor \(j\), and constants and factors \(C_j(\tau)\) and \(D_j(\tau)\)
that again do not vary over time. Then
\(R(t,\tau) = \sum_j [C_j(\tau) + D_j(\tau) r_j(t)]\). Mixtures of
processes like that can give a much greater variety of yield-curve
shapes, especially mixtures of at least three processes. Still the
changes in the yield curves over time come from the short rates
components \(r_j(t)\), as the \(C_j(\tau), D_j(\tau)\) functions still
do not vary over time \(t\).

Interest rates are modeled as continuous processes, using Brownian
motion, but are fit with discrete data, so can be represented by
time-series approximations. (A Brownian motion is a continuous process
whose value at any future time is normally distributed with mean equal
to the current value and variance equal to the time interval.)
Parameters are usually scaled as annual changes, with the rates
incremented in fractions of a year \(dt\). For instance, \(dt\) might be
0.004, for an increment of 1/250th of a year, which is often used to
represent one trading day. The advantage of starting with Brownian
motion is that any increment can be taken as an approximation. We will
start by assuming an unspecified increment \(dt\).

The basic building blocks of affine models are the CIR and Vasicek
processes. These are specified by the distribution of the incremental
change \(dr(t) = r(t+dt) - r(t)\), which is stochastic. More general
affine models are built up as combinations of these two. In both of
these processes, the distribution of the increments is normal, which we
will write as \(\mathcal{N}(\mu,\sigma)\). In the Vasicek process, from
Vasicek (1977):
\[dr(t)\sim \mathcal{N}\left([\omega - \kappa r(t)]dt, \sigma\sqrt{dt}\right)\]where
\(\omega, \kappa, \sigma\) are parameters. The square root of \(dt\) is
taken because the variance is proportional to the time increment \(dt\).
The constants \(C(\tau),D(\tau)\) for each maturity \(\tau\) are
calculated by closed form but slightly complicated formulas.

The CIR process, from Cox, Ingersoll, and Ross (1985), is similar, with
\[dr(t) \sim \mathcal{N} \left([\omega - \kappa r(t)]dt, \sqrt{\beta r(t)dt}\right)\]Its
incremental variance is proportional to its latest value \(r(t)\). The
variance parameter here is denoted as \(\beta\) rather than \(\sigma^2\)
because it is not the variance but a factor which when combined with
\(r(t)\) gives the variance. The processes can be simulated in steps of
any desired (but small) \(dt\) by sampling from these normal
distributions. Also these distributions become the priors for the
incremental changes in the values of the processes at each subsequent
step when doing Bayesian estimation from the historical processes.

The CIR process cannot go negative: it is a continuous process, so if
\(r(t)=0\), the standard deviation also becomes zero, and the process
increments by \(\omega dt\), which is positive. However when simulating
the process, if it gets to a small positive value, the next discrete
simulation can become negative. A work-around derives from the fact that
the constantly changing volatility results in a gamma-like distribution
for the process over time. This is actually the non-central chi-squared
distribution, and for step size \(dt\), the next value of the process
has mean and variance:
\[\mu = r(t)e^{-\kappa dt} + \omega \left(1 - e^{-\kappa dt}\right)/\kappa\]
\[V = \frac{\beta}{2\kappa} \left(1 - e^{-\kappa dt}\right) \left[2r(t)e^{-\kappa dt}+ \omega \left(1 - e^{-\kappa dt}\right)/\kappa \right]\]The
non-central chi-square is close to a gamma distribution, so it can be
approximated by a gamma distribution with this mean and variance. This
way the process would not become negative in simulations.

The long-term variance is \(\beta \omega/{2}\kappa^2\). For the Vasicek
process, this is \(\sigma^2/{2\kappa}\). The Vasicek long-term
distribution is normal with mean \(\omega/\kappa\) and this variance,
while the CIR long-term distribution is gamma. Both the variances follow
the algorithm: expected value of the process change variance divided by
twice the speed of mean reversion. These could be used in the priors for
the first values of each process, in that they have been going on a long
time.

As discussed below, multifactor processes can have a few Vasicek and CIR
processes, or can even include combined forms. The CIR processes must be
independent or positively correlated to be arbitrage-free, but Vasicek
processes can be positively or negatively correlated. Apparently
negatively correlated processes provide a realistic variety of yield
curves.

\textbf{\emph{3a Yield curves and market price of risk}}\\
Dai and Singleton (2000) provide a comprehensive characterization of
affine models. Like Jagannathan, Kaplin, and Sun (2003) also recommend,
they focus on three-factor models. We start with three-factor
combinations of the Vasicek and CIR models that have closed form
\(C_j(\tau)\) and \(D_j(\tau)\) functions.

The yield-curve formulas are derived as expectations of related
risk-neutral processes. The idea of risk-neutral processes is to add
risk load directly into the processes, so the expected value of the bond
value at maturity discounted along the risk-neutral rate process is the
risk-loaded price of the bond. This allows risk loads to be put into
bonds of various maturities in a consistent way. The fundamental theorem
of asset pricing says that arbitrage-free prices must be the expected
values of admissible transformed processes -- i.e., ones that do not
change the set of outcomes that have non-zero probabilities. Such
transforms here change the \(\omega\) and \(\kappa\) parameters, but not
the \(\sigma\) or \(\beta\) parameters.

The simplest transform is called completely affine. Under it, the mean
\([\omega - \kappa r(t)]dt\) of these process changes to
\([\widetilde{\omega} - \widetilde{\kappa} r(t)]dt\). This uses a
market-price-of-risk parameter \(\lambda\), which is estimated when
fitting the model. For the Vasicek process, it works out that
\(\widetilde{\kappa} = \kappa\) and
\(\widetilde{\omega} = \omega - \sigma\lambda\). For the CIR process, on
the other hand, \(\widetilde{\kappa} = \kappa + \beta\lambda\) and
\(\widetilde{\omega} = \omega\). The transformed parameters are called
the risk-neutral parameters. The derivation of the \(C(\tau)\) and
\(D(\tau)\) functions assumes that the bond prices are expected values
under the transformed process.

There is a more general approach to the market price of risk called
essentially affine. It does not affect the previously-transformed CIR
process but for each Vasicek process there is another market price
parameter \(\psi\). Then \[\widetilde{\omega} = \omega - \sigma\lambda\]
\[\widetilde{\kappa} = \kappa + \sigma\psi\]

In either case, the values of \(C(\tau),D(\tau)\) are derived from the
risk-neutral processes. Let
\(q = \widetilde{\omega}/\widetilde{\kappa}\). Then for the Vasicek
process, the constants \(C(\tau),D(\tau)\) are:
\[D(\tau) = \frac{1 - e^{-\widetilde{\kappa}\tau}}{\widetilde{\kappa}\tau}\]
\[C(\tau) =  q -qD(\tau) + \left[ \frac{\sigma}{2\widetilde{\kappa}} \right]^2 \left[\widetilde{\kappa}\tau D(\tau)^2 + 2D(\tau) -2\right] \]The
CIR formulas require two intermediate
values:\[h = \sqrt{\widetilde{\kappa}^2 + 2\beta}\]
\[Q(\tau) = \left[(\widetilde{\kappa}+h)\left(e^{h \tau}-1 \right) +2h \right]^{-1}\]Then:
\[C(\tau) = -\frac{\widetilde{\omega}}{\tau \beta} \left(2log[2hQ(\tau)]+\widetilde{\kappa}\tau+h\tau\right)\]

\[D(\tau) = 2Q(\tau)\left(e^{h\tau} - 1 \right)/\tau\] The \(D\)
functions start at 1 for \(\tau = 0\) and decline from there. That means
that the longer maturities change less as \(r(t)\) changes, and so have
less volatility than do the shorter maturities. This is a pretty
standard property of yield curves but did not hold over much of the last
decade. These models would probably not do very well when the short
rates are so low. Having greater volatility in the shorter rates also
builds in the property that longer spreads get lower when the short rate
rises.

The \(C(\tau),D(\tau)\) functions depend on \(\widetilde{\kappa}\) and
\(\widetilde{\omega}\), so some analysts prefer to make these the
primary parameters to estimate, and then calculate \(\kappa,\omega\) by
reversing the \(\lambda\) and \(\psi\) adjustments. In a sense, the
\(C(\tau)\) and \(D(\tau)\) functions basically define the model, as
they are constant over time while the short-rates evolve stochastically.
Fitting the model then comes down to deriving these functions from the
risk-neutral processes. That is why these processes are often taken as
the starting point. Doing it this way also seems to help numerically
with parameter estimation, as \(\lambda\) and \(\psi\) then do not go
into estimating the functions. That approach seems to speed up the
estimation. However we take it one step further.

For a thee-factor model with one CIR and two Vasiceks, in the completely
affine case, this method produces all the risk-neutral parameters, plus
three real-world parameters: the CIR \(\kappa\) and the two Vasicek
\(\omega\)s. We actually estimate these directly, and the \(\lambda\)s
can be backed out later, if desired. In the essentially affine model it
is similar, but now the real-world and risk-neutral parameters can all
be estimated independently, just with the CIR
\(\omega = \widetilde{\omega}\). This also holds for the more general
models discussed later for the essentially affine risk loads. It is
easier for the software to do the estimation this way, as there are less
interactions among the parameters that are specified in advance.

In multi-factor models, the CIR and Vasicek processes \(r_j\) are taken
as unobserved components of the short rate. Any number of independent
processes can be combined in this way. Then the yield curves from each
model are treated as partial interest rates, and they add up to the
interest rates for each maturity. The \(C_j(\tau)\) functions add over
\(j\), as do the \(D_j(\tau)r_j(t)\) terms.

For two correlated Vasicek processes, the \(D_j(\tau)r_j(t)\) terms add,
but an adjustment to the \(C_j(\tau)\) functions is needed. If the
correlation is \(\rho\), Brigo and Mercurio (2001), p.~135, as well as
Troiani (2017), give the
adjustment:\[C(\tau) = C_1(\tau) + C_2(\tau) + \frac{\rho \sigma_1 \sigma_2}{\widetilde{\kappa}_1\widetilde{\kappa}_2}\left[\frac{e^{-\tau (\widetilde{\kappa}_1+\widetilde{\kappa}_2)} - 1}{\tau (\widetilde{\kappa}_1+\widetilde{\kappa}_2)} +D_1(\tau)+D_2(\tau) - 1 \right]\]If
there are additional independent Vasicek or CIR processes, their \(C\)
functions add in as well, and so do the \(D_j(\tau)r_(t)\) terms.

Bolder (2001) shows the adjustment for any number of correlated
Vasiceks:
\[C(\tau) = \sum_j C_j(\tau) + \sum_{i,j:j\neq i}\frac{\rho_{ij} \sigma_i \sigma_j} {\widetilde{\kappa}_i\widetilde{\kappa}_j} \left[\frac{e^{-\tau (\widetilde{\kappa}_i+\widetilde{\kappa}_j)} - 1}{\tau (\widetilde{\kappa}_i+\widetilde{\kappa}_j)} +D_i(\tau)+D_j(\tau) - 1 \right]\]This
uses the sum of the correlation adjustments across all of the binary
correlations.

Multi-factor models allow a slight generalization to the idea that the
sum of the vector \(r(t)\) of the short-rate components at time \(t\) is
the actual short rate \(r_s(t)\). For instance, three-factor models
allow a constant \(\delta_0 \geq 0\) and a positive three-vector
\(\delta\) so that \(r_s(t) =\delta_0 + \delta'r(t)\). The default
assumption, assumed above, is that \(\delta_0 = 0\) and
\(\delta_j =1, j>0\). We introduce
\(\gamma_j = \delta_j - 1 > -1, j>0\). The result of this is to add
\(\delta_0\) to \(C(\tau)\), and \(\gamma_j\) to \(D_j(\tau)\). Dai and
Singleton (2000) argue that in a three-factor model, nothing is lost by
setting \(\gamma_2 = \gamma_3 =0\). We adopt this approach. Feldhütter
(2016) estimates \(\delta_0\) and \(\gamma_1\) as both less than 0.03.
He leaves \(\gamma_2 , \gamma_3\) in the model, but they both come out
virtually equal to zero.

All of this provides quite a bit that can be done in closed form.
Starting with one CIR and two Vasiceks, you can add correlation to the
Vasiceks, which is one more parameter. The Vasiceks can also use
essentially affine market prices of risk, adding two more parameters.
And two more come from \(\delta_0\) and \(\gamma_1\). This gives quite a
lot of flexibility to the model building. You start off with 3
\(\widetilde{\kappa}\) parameters, 3 \(\widetilde{\omega}\)'s, 2
\(\sigma\)'s and \(\beta\) to make the \(C(\tau),D(\tau)\) functions.
The correlation \(\rho\) affects those, as do \(\delta_0\) and
\(\gamma_1\). Then the market-price-of-risk adjustments add five more
factors to give the real-world yield-curve processes.

The more general affine models, discussed in Appendix 1, require solving
systems of ordinary differential equations for the \(C(\tau),D(\tau)\)
functions. There is convenient software for this, illustrated in the
sample code in Appendix 3, but it does add to model fitting time. The
fairly complicated formulas for \(C,D\) for the CIR and Vasicek models
are actually solutions of these differential equations.

In the models discussed so far, the drifts -- the incremental mean
changes of the processes -- depend on the immediately previous values of
their processes. The more general models allow the immediately previous
values of all the processes to go into the drift of a process. The CIR
variance is a multiple of the process itself. In the more general
models, every process variance is allowed to be a linear function of the
CIR process.

The CIR process is very delicate with respect to any possible
generalization and is usually left the same in the more general models.
Below is an example of the evolution formula for the first Vasicek
process, assumed to be the second process in a model with one CIR and
two Vasicek processes:
\[dr_2(t) \sim \mathcal{N}\left( \left[\omega_2 - \kappa_{21} r_1(t) - \kappa_{22} r_2(t) - \kappa_{23} r_3(t)\right]dt, \sigma_2\sqrt{1+\beta_2r_1(t)dt}\right) \]This
lowers subsequent values of this process when any of the processes is
high, and also allows the stochastic volatility from the first process
to go into each of the process variances. We fit this model below for
comparison to the closed-form models. The other Vasicek process is like
this one, but has an extra volatility term to produce correlation with
this process. The first process is a plain CIR. With essentially affine
market price of risk and the \(\delta_0, \gamma_1\) parameters, this is
the maximal model we fit. Below we call this model 7k3b, because it has
7 \(\kappa\)'s and 3 \(\beta\)'s. In the general form, the \(\kappa\)'s
are in a 3x3 matrix \(K\), but in this CIR,
\(\kappa_{12} = \kappa_{13}=0\).

Appendix 2 goes into the model fitting by MCMC. This requires specifying
postulated (prior) distributions for the parameters, but these can be
changed after seeing the implied conditional distributions of the
parameters given the data. Thus they are not exactly like the Bayesian
view of prior beliefs. Historical values of \(r_j(t)\) are not
parameters of the model, but they are projected as a step in parameter
estimation. These are not allowed to fluctuate freely, however. Their
prior distributions are defined by the process evolution equations. Thus
the next value of a process would be normally distributed according to
the evolution assumptions of the process. (In MCMC you do not have to
specify the form of the posterior (conditional given the data) parameter
distributions -- they are sampled numerically based on the priors, the
likelihood, and the data.) All of this produces projections for the
history of the partial short-rate processes \(r_j\), and so also
provides fitted values of the yield curve at every time point in the
data. More detail is in Appendix 2.

\textbf{4 Results of Fitting}\\
A key issue in fitting models to yield-curve data is the choice of
period and data to use. The models assume zero-coupon bonds, but there
is no raw data on those. Some data series have been constructed, for
instance by the Wall Street Journal. Here we instead use US Treasury
Constant Maturity Rates, available from the St.~Louis Fed FRED database
(\url{https://fred.stlouisfed.org/categories/115}). It would be rare to
have bonds on the market that mature in exactly 5 years, for instance,
and the Fed estimates 5-year, etc., rates by interpolating related
yields on actual trades. These rates assume semi-annual interest
payments at whatever rates the actual bonds carry. The data is
considered to be estimates of the yields, each with a distribution
around the actual rate. We assume these distributions are all normal,
with standard deviations \(=\sigma_y\), which is estimated in the model
fitting using the differences of the fitted and observed rates.

Short-rates have come out of the very-low-yield period following the
economic crash of 2008, with the Fed increasing their rates a fair
amount. Fitting older data would not be representative of the market we
are in now, and even starting at the beginning of the period of rate
increases creates a problem of models with built-in upward trend. We
chose to use week-ending rates from 1/5/2018 to 6/21/2019 for maturities
of 1, 2, 3, 5, 7, 10, 20, and 30 years. Rates shorter than this follow
the Fed much more than the market so are not really generated from the
same process. This is largely true for the one-year rates as well, but
they are too important to leave out. That rate was at 1.82\% at the
beginning of our data, and ended 1.98\%. It got as high as 2.7\% along
the way. All the other rates also went up for much of this period, but
ended up lower than they started. For instance, two-year rates started
at 1.95\%, got as high as 2.94\%, and ended at 1.79\%. Thirty-year rates
went from 2.8\% up to 3.42\% then back down to 2.56\% The Fed Funds Rate
went up for a while then leveled off, not decreasing at the end. The
one-year rate seems more influenced by this, and ended higher than the
two, three and five-year rates.

We fit four models to that data. VVV is a closed-form model consisting
of three correlated Vasicek processes, with essentially affine market
prices of risk. This is considered an \(A_0(3)\) model as there are
three processes and none of them affect the volatility of the rates. The
other three models are \(A_1(3)\) models, having a single CIR process.
CVV is a completely affine model with one CIR and two correlated Vasicek
processes. CVV+ is our maximal closed-form model, and is like CVV but is
essentially affine and includes parameters for \(\delta_0\) and
\(\gamma_1\). 7k3b is the overall maximal ODE model that we fit. It has
7 \(\kappa\) parameters -- all but \(\kappa_{12},\kappa_{13}\) -- and
all three \(\beta\) parameters. The CVV model has 14 parameters -- 3 for
each risk-neutral process, 3 additional real-world parameters, \(\rho\),
and \(\sigma_y\). VVV has 19 parameters, so 5 more: 2 more correlations,
and 6 real-world parameters -- a \(\kappa\) and a \(\omega\) for each
process. CVV+ has 18 parameters. It only has 1 correlation, but has
\(\delta_0\) and \(\gamma_1\) as well as 5 real-world parameters, as
\(\widetilde{\omega}_1 = \omega_1\). 7k3b actually has 10 more
parameters than this, so 28 in total. Compared to CVV+ it has 4 more
\(\widetilde{\kappa}\)s, 2 more \(\beta\)s, and 4 more \(\kappa\)s.

Table 7 shows the fitted parameters for each model, along with:
\(\sigma_y\); the implied real-world model means \(\mu_j\) for each
process and their implied total short rate; the loo penalized
loglikelihood goodness-of-fit measure for MCMC fits; the loo penalty for
parameters (difference from LL); and the implied loglikelihood.

\begin{table}[]
\centering
\caption{Fitted Models}
\begin{tabular}{crrrr}
Variable                  & VVV    & CVV    & CVV+    & 7k3b   \\
$\widetilde{\omega}_1$    & -30.3  & 1.33   & 1.19    & 0.113  \\
$\widetilde{\omega}_2$    & -0.627 & -62    & -19.28  & -2.758 \\
$\widetilde{\omega}_3$    & 30.3   & 62.2   & 19.91   & 1.649  \\
$\widetilde{\kappa}_{11}$ & 0.07   & 0.027  & 0.165   & 0.007  \\
$\widetilde{\kappa}_{22}$ & 1.076  & 0.063  & 0.07    & 0.330  \\
$\widetilde{\kappa}_{33}$ & 0.055  & 0.052  & 0.047   & 0.224  \\
$\widetilde{\kappa}_{21}$ & -      & -      & -       & 0.047  \\
$\widetilde{\kappa}_{23}$ & -      & -      & -       & 0.122  \\
$\widetilde{\kappa}_{31}$ & -      & -      & -       & -0.708 \\
$\widetilde{\kappa}_{32}$ & -      & -      & -       & -0.933 \\
$\sigma_1$                & 1.64   & 1      & 1       & 1      \\
$\sigma_2$                & 0.83   & 2.58   & 1.17    & 0.159  \\
$\sigma_3$                & 1.92   & 2.9    & 1.31    & 0.071  \\
$\beta_1$                 & -      & 1.85   & 1.19    & 9E-05  \\
$\beta_2$                 & -      & -      & -       & 1.078  \\
$\beta_3$                 & -      & -      & -       & 2.703  \\
$\omega_1$                & 24.1   & 1.33   & 1.19    & 0.113  \\
$\omega_2$                & -0.296 & -0.596 & 7.045   & -6.626 \\
$\omega_3$                & -19.2  & 0.43   & -22.272 & 2.908  \\
$\kappa_{11}$             & 0.972  & 2.432  & 1.8667  & 0.012  \\
$\kappa_{22}$             & 1.906  & -      & 0.7369  & 0.918  \\
$\kappa_{33}$             & 1.0534 & -      & 2.4443  & 1.094  \\
$\kappa_{21}$             & -      & -      & -       & 0.023  \\
$\kappa_{23}$             & -      & -      & -       & -0.678 \\
$\kappa_{31}$             & -      & -      & -       & 0.133  \\
$\kappa_{32}$             & -      & -      & -       & -0.228 \\
$\rho_{12}$               & 0.17   & -      & -       & -      \\
$\rho_{13}$               & -0.93  & -      & -       & -      \\
$\rho_{23}$               & -0.47  & -0.96  & -0.81   & -0.65  \\
$\delta_0$                & -      & -      & 0.021   & 0.253  \\
$\gamma_1$                & -      & -      & -0.032  & 0.001  \\
$\sigma_y$                & 0.025  & 0.023  & 0.023   & 0.019  \\
$\mu_1$                   & 9.41   & 0.55   & 0.64    & 9.52   \\
$\mu_2$                   & -0.46  & -13.05 & 9.36    & -5.7   \\
$\mu_3$                   & -6.35  & 6.2    & -9.09   & -0.05  \\
Mean short rate           & 2.6    & -6.3   & 0.91    & 4.02   \\
loo                       & 1312.9 & 1324.7 & 1333.3  & 1462.1 \\
penalty                   & 134    & 175.7  & 184.2   & 183.5  \\
LL                        & 1446.9 & 1500.4 & 1517.5  & 1645.6
\end{tabular}
\end{table}

The VVV model does not fit as well as any of the CIR models, according
to loo. The Vasicek and CIR yield curves can have somewhat different
shapes, which could be contributing to this. Also stochastic volatility
may be a feature of the data, and Vasicek models do not capture that.
The essentially affine version of the CVV model, including the
\(\delta_0,\gamma_1\) adjustments to the \(C(\tau),D(\tau)\) functions,
is better-fitting than the basic version. The full model, even when
considering all the additional parameters, is better yet, both in loo
and the residual standard deviation \(\sigma_y\). The 10 additional
parameters make the model much more flexible, and apparently this data
can make use of that flexibility.

All of the projected short-rates (here for 3 processes and 77 data
points, so 231 in total) are considered parameters for testing goodness
of fit, in addition to the model parameters and the market prices of
risk. Thus there are between 245 and 259 parameters that go into the
parameter penalty. The penalties are much less than this, and are not
always higher with more parameters.

The loo penalty is not calculated as a multiple of the number of
parameters but rather comes from a cross-validation approach. The
penalized loglikelihood is an estimate of the loglikelihood (LL) of the
fitted model for a new independent sample. The penalty is how much lower
the penalized likelihood is from the actual LL. It is an estimate of the
sample bias, which tends to increase for more parameters, but not in a
readily-predictable way for non-linear models like these. The penalties
here are less than the actual number of parameters partly because the
estimated historical short-rates are highly constrained and do not act
like independent parameters.

Ye (1998) defines the generalized degrees of freedom used by a data
point in a nonlinear model as the derivative of the fitted point wrt the
actual data point. These derivatives constitute the diagonal of the hat
matrix in linear models. The sum of the resulting dofs is the effective
number of parameters, and these can then be penalized by AIC, etc.
Constrained parameters do not give the data points much power to move
the fitted values towards them, so the derivatives and the effective
number of parameters are reduced. Something similar happens in loo. The
loglikelihood at a point is penalized by how much it would be reduced if
that point were left out of the sample. Again if the parameters are
highly constrained, the loglikelihood at a point is not affected much by
leaving it out of the estimation.

\begin{figure}
\centering
\includegraphics{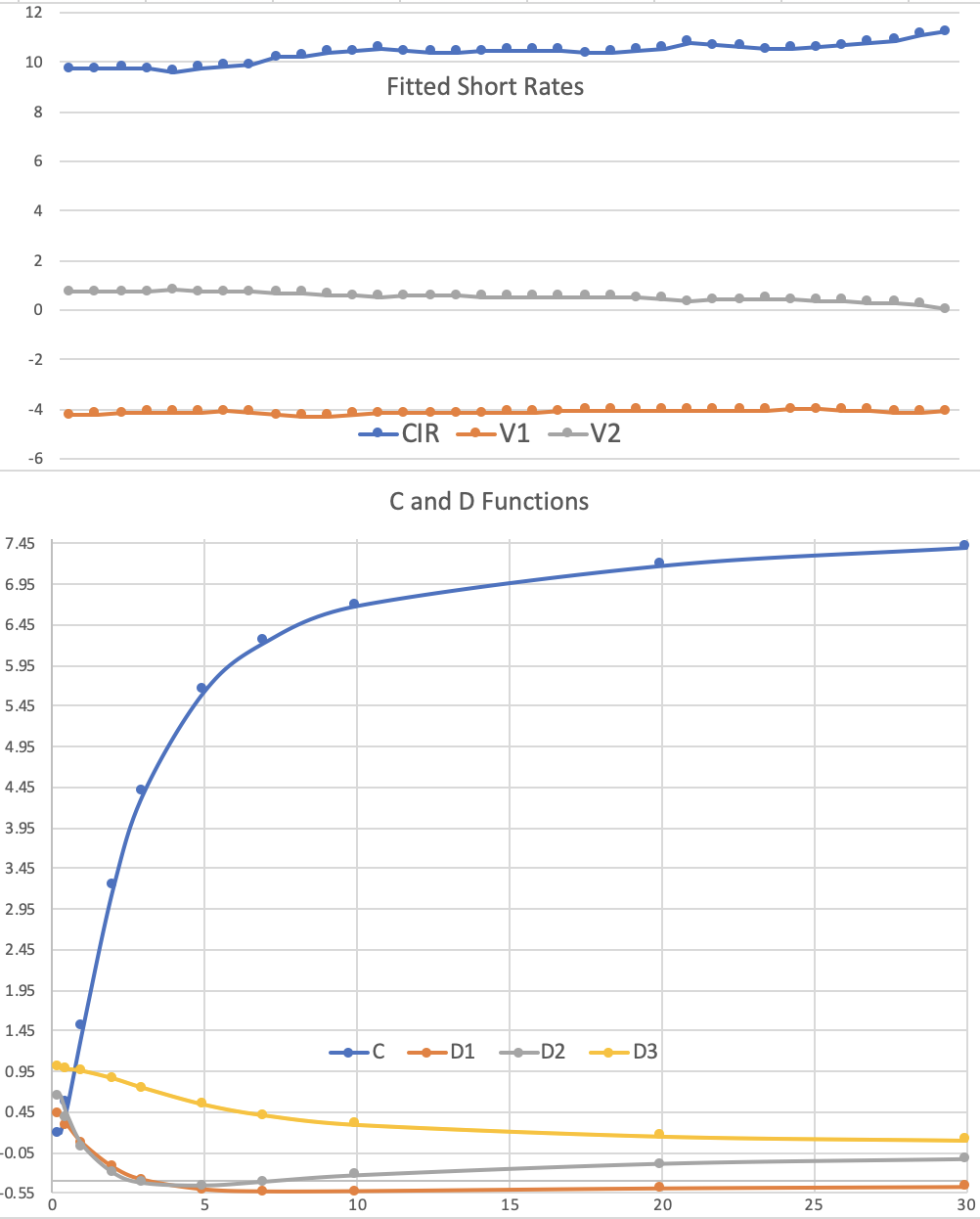}
\caption{Projected Historical Short Rates and C, D Functions
\label{fig1}}
\end{figure}

\begin{figure}
\centering
\includegraphics{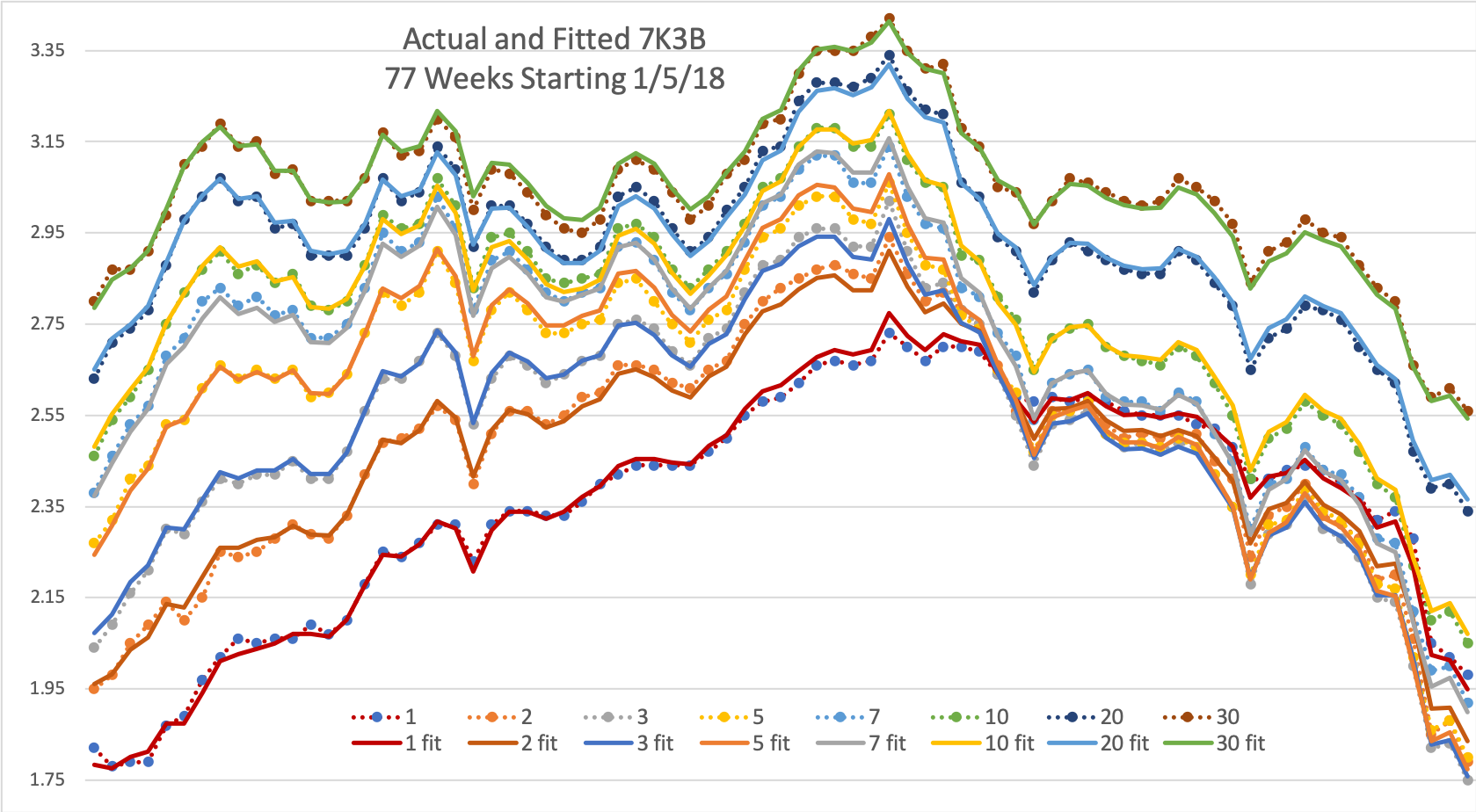}
\caption{Actual vs.~Fitted Yield Curves \label{fig1}}
\end{figure}

Figures 10 and 11 show the fitted short rates, the \(C(\tau),D(\tau)\)
functions, and the resulting rate fits for the 7k3b model.

\textbf{5 Applying Tests}\\
We now apply several of the tests from Section 2 to the VVV and 7k3b
models for illustration. We first extracted the sampled parameter
distributions from the Stan output for each model. Then we simulated two
years of future rates for the two models, and these simulated yield
curves were the used for the tests. This is similar to the common
situation of the insurer only having the generated scenario sets and not
the model fitting comparisons.

The simulations started from the sampled parameter distributions, which
is a way of including parameter uncertainty. Each simulation starts by
drawing a parameter sample at random. Then the ending values of the
three processes in that sample are the starting previous values for
their first simulations. The simulations are done by computing the drift
and the variance using the real-world processes, then doing a random
draw for the next value of each process. The \(C\) and \(D\) functions
(already in the parameter samples) are based on the risk-neutral
processes and are not simulated except as implied by the draw of the
sample. We simulate at monthly intervals for two years for each process,
and the two year-ending yield curves are used in the tests.

Table 8 shows the moments of the simulated rates for the two models by
year. The mean yield curves are basically upward for the VVV model,
while the 7k3b model projects the current curve shape, which is high on
both ends and low in the middle, which seems better. The standard
deviations generally decline with maturity, which is probably all you
can look for there at this point, as the right shape of the volatilities
by maturity is still emerging. The skewness is moderately negative for
the VVV model, indicating more downward risk in yields. It is also
negative for the longer rates in the 7k3b model, but is positive for the
shorter rates. That could be coming from the recent experience, which
had movements like that.

\begin{table}[]
\centering
\caption{Scenario Moments}
\begin{tabular}{lrrrrrrrr}
Maturity    & 1     & 2     & 3     & 5     & 7     & 10    & 20    & 30    \\
VVV Year 1  &       &       &       &       &       &       &       &       \\
mean        & 1.90  & 1.87  & 1.89  & 1.97  & 2.06  & 2.18  & 2.48  & 2.70  \\
std         & 0.40  & 0.44  & 0.46  & 0.47  & 0.47  & 0.45  & 0.38  & 0.32  \\
skw         & -0.55 & -0.50 & -0.45 & -0.41 & -0.39 & -0.37 & -0.35 & -0.34 \\
VVV Year 2  &       &       &       &       &       &       &       &       \\
mean        & 1.94  & 1.92  & 1.95  & 2.03  & 2.12  & 2.24  & 2.52  & 2.73  \\
std         & 0.53  & 0.56  & 0.58  & 0.58  & 0.57  & 0.54  & 0.45  & 0.38  \\
skw         & -0.86 & -0.77 & -0.71 & -0.64 & -0.61 & -0.59 & -0.55 & -0.53 \\
7k3b Year 1 &       &       &       &       &       &       &       &       \\
mean    & 2.74  & 2.56  & 2.45  & 2.38  & 2.42  & 2.51  & 2.71  & 2.86  \\
std & 0.82  & 0.76  & 0.76  & 0.73  & 0.68  & 0.60  & 0.45  & 0.40  \\
skw & 0.18  & 0.04  & -0.08 & -0.23 & -0.28 & -0.30 & -0.30 & -0.30  \\
7k3b Year 2 &       &       &       &       &       &       &       &       \\
mean    & 3.29  & 2.85  & 2.57  & 2.32  & 2.30  & 2.39  & 2.65  & 2.82  \\
std & 1.47  & 1.01  & 0.90  & 0.93  & 0.92  & 0.83  & 0.59  & 0.49  \\
skw & 0.94  & 0.68  & 0.24  & -0.85 & -1.20 & -1.27 & -1.02 & -0.78       
\end{tabular}
\end{table}

Table 9 has the spread regressions for the 3-year -- 30-year spreads as
a function of the 1-year rates and the Campbell-Schiller regressions for
2 and 3-year rates. The spread regressions' slopes and standard errors
are compatible with recent history for 7k3g but are pretty small for
VVV. The expected slopes for the Campbell-Schiller regressions are
negative. That is all consistent with the targets for these tests, and
what we would want for the exact magnitudes of the slopes is not clear
beyond those signs.

\begin{table}[]
\centering
\caption{Scenario Regressions}
\begin{tabular}{lrrr}
                                    & Spread Slope & Spread SE & Campbell-Schiller Slope \\
VVV Year 1 (2 Year $\tau$ for C-S)  & -0.42        & 0.01      & -0.99                   \\
VVV Year 2 (3 Year $\tau$ for C-S)  & -0.29        & 0.01      & -1.63                   \\
7k3b Year 1 (2 Year $\tau$ for C-S) & -0.61        & 0.06      & -2.63                   \\
7k3b Year 2 (3 Year $\tau$ for C-S) & -0.45        & 0.09      & -1.99                  
\end{tabular}
\end{table}

We did principal component analysis on the actual data, the fitted means
for all four models, and the two years of simulated scenarios for the
VVV and 7k3b models. For this we used the R function prcomp, which takes
as input an array, with the variables in the columns and the observation
times in rows. The first three components explain the vast majority of
the variation in the data. An indicator of the degree of complexity of
the yield curves is provided by the percentage of variance explained by
the third principal component. This is small for the short time period
we have, but was greatest in the data itself, and was higher for the
fitted values than for the simulated scenarios. The VVV model looks too
weak in the third PC's proportion of variance in the simulations. Table
10 shows the results.

\begin{table}[]
\centering
\caption{PCA Comparisons}
\begin{tabular}{lrrr}
data                   & PC1   & PC2   & PC3   \\
Standard deviation     & 0.657 & 0.254 & 0.083 \\
Proportion of Variance & 0.857 & 0.128 & 0.014 \\
Cumulative Proportion  & 0.857 & 0.985 & 0.999 \\
                       &       &       &       \\
fit vvv                & PC1   & PC2   & PC3   \\
Standard deviation     & 0.655 & 0.251 & 0.070 \\
Proportion of Variance & 0.863 & 0.127 & 0.010 \\
Cumulative Proportion  & 0.863 & 0.990 & 1.000 \\
                       &       &       &       \\
fit cvv                & PC1   & PC2   & PC3   \\
Standard deviation     & 0.656 & 0.252 & 0.074 \\
Proportion of Variance & 0.862 & 0.127 & 0.011 \\
Cumulative Proportion  & 0.862 & 0.989 & 1.000 \\
                       &       &       &       \\
fit cvv+               & PC1   & PC2   & PC3   \\
Standard deviation     & 0.655 & 0.253 & 0.074 \\
Proportion of Variance & 0.861 & 0.128 & 0.011 \\
Cumulative Proportion  & 0.861 & 0.989 & 1.000 \\
                       &       &       &       \\
fit 7k                 & PC1   & PC2   & PC3   \\
Standard deviation     & 0.657 & 0.254 & 0.079 \\
Proportion of Variance & 0.859 & 0.128 & 0.013 \\
Cumulative Proportion  & 0.859 & 0.987 & 1.000 \\
                       &       &       &       \\
vvv yr1                & PC1   & PC2   & PC3   \\
Standard deviation     & 1.189 & 0.187 & 0.062 \\
Proportion of Variance & 0.973 & 0.024 & 0.003 \\
Cumulative Proportion  & 0.973 & 0.997 & 1.000 \\
                       &       &       &       \\
vvv yr2                & PC1   & PC2   & PC3   \\
Standard deviation     & 1.481 & 0.211 & 0.072 \\
Proportion of Variance & 0.978 & 0.020 & 0.002 \\
Cumulative Proportion  & 0.978 & 0.998 & 1.000 \\
                       &       &       &       \\
7k yr1                 & PC1   & PC2   & PC3   \\
Standard deviation     & 3.067 & 0.749 & 0.165 \\
Proportion of Variance & 0.789 & 0.203 & 0.009 \\
Cumulative Proportion  & 0.789 & 0.991 & 1.000 \\
                       &       &       &       \\
7k yr2                 & PC1   & PC2   & PC3   \\
Standard deviation     & 1.975 & 1.734 & 0.222 \\
Proportion of Variance & 0.561 & 0.432 & 0.007 \\
Cumulative Proportion  & 0.561 & 0.993 & 1.000
\end{tabular}
\end{table}

\textbf{Conclusion}\\
The last 30 or 40 years of yield curve history, instead of being the new
normal, are starting to look more like a one-time aberration, driven by
the post-WWII financing needs. We are now seeing interest rates more
like the 1870 -- 1970 period, where the long rates rarely got above 4\%.
This makes constructing tests based on yield-curve standard behavior
challenging.

We started by reviewing the tests that had been used historically in the
related literature. Most of them are still useful with updated targets,
although these are fairly broad given the uncertainty about the
properties of interest rates in the emerging era. We applied the tests
to a few models fit to recent data, and the models did reasonably well
in the context of these broad guidelines, with the better-fitting model
also doing better on the tests.

The models used were affine models, which build up the yield curves as
weighted sums of partial short-rate processes. The weights for this are
time invariant but are a bit complicated to calculate, and for the most
general models require numerical solutions of ordinary differential
equations. Several closed form models where the differential equations
have been solved in closed form provide reasonable fits, but not as good
as the more general models.

MCMC estimation provides a direct way to fit the models, and produces
fitted values for the historical processes.

\newpage

\textbf{Appendix 1 -- General Affine Models}

\textbf{\emph{A1a Matrix Notation}}\\
We follow the notation of Dai and Singleton (2000) for affine models.
The model combining two Vasiceks and one CIR is classified as an
\(A_1(3)\) model. That means that there are three processes, with one of
them affecting the variance -- in this case the CIR process. For
consistency with more general models, we replace \(\kappa_j\) with
\(\kappa_{jj}\) and \(\sigma_j\) with \(\sigma_{jj}\). Then our model,
with the CIR model first, can be
written:\[dr_1(t) \sim \mathcal{N} \left([\omega_1 - \kappa_{11}r_1(t)]dt, \sqrt{\beta_1r_1(t)dt}\right)\]
\[dr_2(t) \sim \mathcal{N} \left([\omega_2 - \kappa_{22} r_2(t)]dt , \sigma_{22}\sqrt{dt}\right)\]
\[dr_3(t) \sim \mathcal{N} \left([\omega_3 - \kappa_{33} r_3(t)]dt , \sigma_{33}\sqrt{dt}\right)\]
\[corr\left(dr_2(t),dr_3(t)\right)=\rho\]While this is a convenient way
to show the correlation, more calculation detail is needed for use as a
prior or for simulation. The bivariate normal prior can be used for the
two Vasicek processes, specifying their covariance matrix as:
\[Cov(dr_2(t),dr_3(t))=\left(\begin{array}{cc} 
\sigma_{22}^2 & \rho\sigma_{22}\sigma_{33} \\
\rho\sigma_{22}\sigma_{33} & \sigma_{33}^2
\end{array}\right)dt\]We do that in the code. For simulation, it is
useful to be able to show these evolution equations in terms of
independent standard normal draws. For this we write
\(dr_j(t) = \mu_j(t)dt + z_j(t)\sqrt{dt}\), where \(z_j(t)dt\) is a mean
zero normal variable. Also we define \(\epsilon_j(t)\) to be a standard
normal random variable observed at time \(t\). Then the vector
\(z(t)dt\) that combines the three processes can be expressed as:
\[z(t)\sqrt{dt} = \left(\begin{array}{ccc} 
\sqrt{\beta_1r_1(t)} & 0 & 0\\
0 & \sigma_{22} & 0\\
0 &\rho \sigma_{22} & \sqrt{1-\rho^2}\sigma_{33}
\end{array}\right)\left(\begin{array}{c} 
\epsilon_1(t) \\ 
\epsilon_2(t) \\
\epsilon_3(t) 
\end{array}\right)\sqrt{dt}\]

This now can be set up in matrix notation as the evolution of a vector
of three processes. Let \(r(t)\) be a column vector of the three
processes, with \(dr(t) \sim \mu(t)dt +z(t)\sqrt{dt}\), and vector of
standard normals \(\epsilon(t)\). Then the general \(A_1(3)\) model
diffusion can be expressed by:\[\mu(t)dt = [\Omega - Kr(t)]dt\]
\[z(t)\sqrt{dt} = \Sigma D(t)\epsilon(t)\sqrt{dt}\]where: \(K\) is the
3x3 matrix of \(\kappa_{i,j}\), with \(\kappa_{12}=\kappa_{13} = 0\);
\(\Omega\) is a 3-vector; \(\Sigma\) is the matrix of \(\sigma_{i,j}\)
values, with \(\sigma_{11}=1,\sigma_{12}=\sigma_{13}=0\) and \(\rho\)
combined into the \(\sigma\) coefficients; \(D(t)\) is a diagonal matrix
with the three values
\(\sqrt{\beta_1 r_1(t)},\sqrt{\alpha_2+\beta_2 r_1(t)},\sqrt{\alpha_3+\beta_3 r_1(t)}\),
with all \(\beta_j\) non-negative and \(\beta_1\) positive. This allows
the CIR process to also contribute to the variances of the other
processes. \(D^2\) is a variance factor, scaled by \(\Sigma^2\). We are
going to assume that \(\alpha_2 = \alpha_3 =1\) in the fitting, as the
\(\sigma\)'s can pick up a lot of the \(\alpha\) effects.

Such general \(A_1(3)\) models provide for interaction among the
processes. For the two Vasicek processes, \(\sigma_{32}\) takes the
place of \(\rho\sigma_{22}\). We set \(\sigma_{23}=0\) for
identifiability, but this is not always done. The off-diagonal
\(\kappa\) terms make a Vasicek process's mean movements responsive to
the current levels of both other processes, as well as its own. That
cannot be done for the CIR process, as it can easily lose its
arbitrage-free property with any adjustments.

Some constraints on the parameters are needed for identifiability. Dai
and Singleton (2000) set up a canonical form of the model that builds in
numerous constraints, but it is not clear how our original three-factor
model would be expressed in it, so we rely instead on the use of priors
on the parameters to get unique parameters. The yield curves are built
up from the short rates using market prices of risk. The CIR and Vasicek
market prices of risk discussed previously are special cases. Here we
follow Christensen (2015) who clarifies the notation and provides
examples.

\textbf{\emph{A1b Market Price of Risk and Bond Yields}}\\
The basic form of market price of risk is the completely affine risk
premium. It uses a 3-vector \(\Lambda(t) = D(t)\lambda\), where
\(\lambda\) is a 3-vector of constants. The risk-neutral process is
produced by a change in the drift term. \(\Sigma D(t) \Lambda(t)dt\) is
subtracted from the drift \([\Omega - Kr(t)]dt\). Some algebra on this
can then be used to express the risk-neutral drift term as
\([\widetilde{\Omega} - \widetilde{K}r(t)]dt\), and this is used to
compute the yield curve, as it was in the three-factor model above.

The term subtracted is similar to the stochastic term of the process. We
need to compute:
\[\widetilde{\Omega}dt - \widetilde{K} r(t)dt = \Omega dt - Kr(t)dt -  \Sigma D(t) \Lambda(t)dt\]The
right side of this is already the risk-neutral drift, but the formulas
for the yield curve use the notation of the left side, so we have to put
it in that form. For the completely affine case,
\(\Sigma D(t) \Lambda(t)dt = \Sigma D(t)^2\lambda\), where \(D(t)^2\) is
diagonal with elements \(\alpha_j + \beta_jr_1(t)\). Then
\(D(t)^2\lambda\) is the vector with elements
\(\lambda_j[\alpha_j + \beta_jr_1(t)]\), which means that
\(\Sigma D(t) \Lambda(t)\) is a 3-vector, but with \(\alpha_1=0\). Each
element is a sum of multiples of \(\alpha_j\) and \(\beta_jr_1(t)\)
terms. The terms with \(\alpha_j\)'s are subtracted from the
\(\omega_j\)'s terms, and this gives \(\widetilde{\Omega}\). All the
factors with \(\beta_jr_1(t)\)'s are subtracted from the drift, so the
coefficients of \(r_j(t)\) are added to \(K\) to produce
\(\widetilde{K}\). Also note that the stochastic part of the process
does not change in the risk-neutral process.

Dai and Singleton (2000) solve some of this for any \(A_1(3)\) model.
Let \(L\) be the diagonal matrix with elements \(\lambda_j\beta_j\) and
\(H\) be the vector of \(\lambda_j\alpha_j\) . Then
\(\widetilde{K} = K + \Sigma L\), and
\(\widetilde{\Omega} = \Omega - \Sigma H\). In practice we reverse this
for estimation efficiency, so we start with the risk-neutral process and
get the real-world process by \(K = \widetilde{K} - \Sigma L\), and
\(\Omega = \widetilde{\Omega} + \Sigma H\).

Above we used formulas from Brigo and Mercurio (2001) to compute the
\(C(\tau)\) and \(D_j(\tau)\) functions in the correlated Vasicek case.
Troiani (2017) illustrate these formulas for the essentially affine
case. For more general models, \(C\) and \(D_j\) functions are not
closed form and need to be computed numerically, from systems of
ordinary differential equations (ODEs). Fast software for solving ODEs
is widely available. The closed-form calculation is considerably faster,
but solving the system numerically is feasible on personal computers.
That's why these functions are considered to be ``almost closed form.''

The system for the \(A_1(3)\) models is expressed in terms of functions
\(A(\tau)\) and \(B_j(\tau)\), with \(C(\tau) = -A(\tau)/\tau\) and
\(D_j(\tau) = B_j(\tau)/\tau\). \(B\) is the vector of the \(B_j\)'s.
Let \(Q(\tau) = \Sigma'B(\tau)B(\tau)'\Sigma\), which is 3x3 and can be
considered to be the square of the three-vector \(\Sigma'B(\tau)\). The
\(j\)th element on the diagonal of \(Q\) is the square of the \(j\)th
element of \(\Sigma'B(\tau)\), so \(Q_{jj} = ([\Sigma'B(\tau]_j)^2\).
Also let \(\beta0_j\) be the vector consisting of \(\beta_j\) followed
by two zeros. The equations are then:
\[\frac{dA(\tau)}{d\tau} = -\widetilde{\Omega}'B(\tau) + \frac{1}{2}\sum_{j=1}^3 ([\Sigma'B(\tau)]_j)^2\alpha_j\]
\[\frac{dB(\tau)}{d\tau} = \textbf{1} -\widetilde{K}'B(\tau) - \frac{1}{2}\sum_{j=1}^3 ([\Sigma'B(\tau)]_j)^2\beta0_j\]Here
\(\textbf{1}\) is a vector of 1's. The starting values are
\(A(0)=B_j(0) = 0\).

We have been assuming that the sum of the vector \(r(t)\) of the three
short-rate components is the actual short rate \(r_s(t)\). But affine
models allow a slight generalization of that with a constant
\(\delta_0\) and a three-vector \(\delta\) so that
\(r_s(t) =\delta_0 + \delta'r(t)\). The differential equations above are
based on the default assumptions that \(\delta_0 = 0\) and
\(\delta = \textbf{1}\). We set \(\gamma = \delta - \textbf{1}\). Then
the differential equations become:
\[\frac{dA(\tau)}{d\tau} = -\widetilde{\Omega}'B(\tau) + \frac{1}{2}\sum_{j=1}^3 ([\Sigma'B(\tau)]_j)^2\alpha_j - \delta_0\]
\[\frac{dB(\tau)}{d\tau} = \gamma + \textbf{1} -\widetilde{K}'B(\tau) - \frac{1}{2}\sum_{j=1}^3 ([\Sigma'B(\tau)]_j)^2\beta0_j\]This
ends up subtracting \(\delta_0\tau\) from \(A(\tau)\) and adding
\(\gamma \tau\) to \(B(\tau)\). Thus \(C(\tau)\) is increased by
\(\delta_0\) and \(D_j(\tau)\) is increased by \(\gamma_j\). Dai and
Singleton (2000) assume \(\delta_0 \geq 0\) and
\(\gamma_2 = \gamma_3 =0\). Feldhütter (2016) estimates \(\delta_0\) and
\(\gamma_1\) as both less than 0.03. These same adjustments to \(C\) and
\(D\) can be done for the closed form cases as well, as discussed
before.

\textbf{\emph{A1c Essentially Affine Market Price of Risk}}\\
Analysts have generally concluded that completely affine risk premiums
are overly restrictive on yield-curve dynamics. Recall that these
subtract \(\Sigma D(t) \Lambda(t)dt\) from the drift
\([\Omega - Kr(t)]dt\), where \(D(t) \Lambda(t)\) is set to
\(D(t)^2 \lambda\), with \(\lambda\) a 3-vector. Greater flexibility is
provided by the essentially affine risk premium, with
\(D(t) \Lambda(t) = D(t)^2 \lambda + J\Psi r(t)\). Here \(J\) is a
diagonal matrix with elements \(0,\alpha_2^{-1/2},\alpha_3^{-1/2}\), and
\(\Psi\) is a 3x3 matrix with the first row all zeros. This makes
\(D(t)\Lambda(t)=\) \[
\left(\begin{array}{ccc} 
\beta_1r_1(t) & 0 & 0\\
0 & \alpha_2+\beta_2r_1(t) & 0\\
0 & 0 & \alpha_3+\beta_3r_1(t)
\end{array}\right)
\left(\begin{array}{c} 
\lambda_1 \\ 
\lambda_2 \\
\lambda_3
\end{array}\right)
+
\left(\begin{array}{ccc} 
0 & 0 & 0\\
0 & \alpha_2^{-1/2} & 0\\
0 & 0 & \alpha_3^{-1/2}
\end{array}\right)
\left(\begin{array}{ccc} 
0 & 0 & 0\\
\psi_{21} & \psi_{22} & \psi_{23}\\
\psi_{31} & \psi_{32} & \psi_{33}
\end{array}\right)
\left(\begin{array}{c} 
r_1(t) \\ 
r_2(t) \\
r_3(t)
\end{array}\right)\]

Multiplying by \(\Sigma\) shows that
\((0,\sigma_{22}\lambda_2\alpha_2,\sigma_{33}\lambda_3\alpha_3)\)' is
subtracted from \(\Omega\), all of the other pieces are added to \(K\).
Alternatively, starting with the risk-neutral coefficients can produce
the real-world coefficients.

\textbf{\emph{A1d Our maximal model}}\\
The maximal model that we fit, which we call 7k3b, is the full general
model with the following restrictions:

\begin{itemize}
    \item $\Sigma$ is diagonal except for $\sigma_{32} = \rho \sigma_{22}$, with $\sigma_{33} = \sqrt{1-\rho^2}\sigma_3$, where $\sigma_3$ is the standard deviation of the original correlated second Vasicek process.
    \item $\sigma_{11} = 1$
    \item $\kappa_{12} = \kappa_{13}  =0$
    \item $\alpha_2 = \alpha_3 = 1$
\end{itemize}

This means that in the model \(dr(t) \sim \mu(t) dt + z(t)\sqrt{dt}\)
\[ \mu(t) = 
\left(\begin{array}{c} 
\omega_1 \\ 
\omega_2 \\
\omega_3
\end{array}\right)-
\left(\begin{array}{ccc} 
\kappa_{11} & 0 & 0\\
\kappa_{21} & \kappa_{22} & \kappa_{23}\\
\kappa_{31} & \kappa_{32} & \kappa_{33}
\end{array}\right)
r(t)\] \[ z(t) = \left(\begin{array}{ccc} 
1 & 0 & 0\\
0 & \sigma_{22} & 0\\
0 & \sigma_{32} & \sigma_{33}
\end{array}\right)
\left(\begin{array}{c} 
\sqrt{\beta_1r_1(t)}\epsilon_1(t)\\
\sqrt{1+\beta_2r_1(t)}\epsilon_2(t)\\
\sqrt{1+\beta_3r_1(t)}\epsilon_3(t)
\end{array}\right) = 
\left(\begin{array}{c} 
\sqrt{\beta_1r_1(t)}\epsilon_1(t)\\
\sigma_{22}\sqrt{1+\beta_2r_1(t)}\epsilon_2(t)\\
\sigma_{32}\sqrt{1+\beta_2r_1(t)}\epsilon_2(t)  + \sigma_{33}\sqrt{1+\beta_3r_1(t)}\epsilon_3(t)
\end{array}\right)  \]

There are thus 7 \(\kappa\)s and 3 \(\beta\)s. The variances of these
processes over a short time \(dt\) are: \(\beta_1r_1(t)\),
\(\sigma_{22}^2 ( \beta_2 r_1(t)+1 ) dt\),
\([ \sigma_{32}^2 ( \beta_2 r_1(t)+1 ) + \sigma_{33}^2 ( \beta_3 r_1(t)+1 ) ]dt\).
The last is from the sum of two normal distributions. The covariance of
the two Vasicek processes is
\(\sigma_{22} \sigma_{32} (1+\beta_2r_1(t))dt\).

The risk-adjustment for the essentially affine version for this model
comes from \[D(t)\Lambda(t)=
\left(\begin{array}{c} 
\lambda_1 \beta_1r_1(t) \\ 
\lambda_2 + \lambda_2 \beta_2r_1(t)  \\
\lambda_3+ \lambda_3 \beta_3r_1(t) 
\end{array}\right)
+
\left(\begin{array}{c} 
0 \\
\psi_{21}r_1(t) + \psi_{22}r_2(t) + \psi_{23}r_3(t)\\
\psi_{31}r_1(t) + \psi_{32}r_2(t) +  \psi_{33}r_3(t)
\end{array}\right)
\]Then \[\Sigma D(t)\Lambda(t)=
\left(\begin{array}{c} 
\lambda_1 \beta_1r_1(t) \\ 
\sigma_{22}\lambda_2 + \sigma_{22}\lambda_2 \beta_2r_1(t)  \\
\sigma_{32}\lambda_2 + \sigma_{32}\lambda_2 \beta_2r_1(t) + \sigma_{33}\lambda_3 + \sigma_{33}\lambda_3 \beta_3r_1(t) 
\end{array}\right)
\] \[
\left(\begin{array}{c} 
0 \\
\sigma_{22}\psi_{21}r_1(t) + \sigma_{22}\psi_{22}r_2(t) + \sigma_{22}\psi_{23}r_3(t)\\
\sigma_{32}\psi_{21}r_1(t) + \sigma_{32}\psi_{22}r_2(t) + \sigma_{32}\psi_{23}r_3(t) + \sigma_{33}\psi_{31}r_1(t) + \sigma_{33}\psi_{32}r_2(t) +  \sigma_{33}\psi_{33}r_3(t)
\end{array}\right)
\]From above, the \(r_2\) mean for this model is:
\[\mu_2(t) = \omega_2 - \kappa_{21} r_1(t) - \kappa_{22} r_2(t) - \kappa_{23} r_3(t) \]
Starting with the risk-neutral version, we can see that
\(\omega_2 = \widetilde{\omega}_2 + \sigma_{22}\lambda_2\). Also,
\(\kappa_{21} = \widetilde{\kappa}_{21} - \sigma_{22}(\lambda_2\beta_2 + \psi_{21})\),
from the middle row of \(z(t)\). The other coefficients are calculated
similarly.

As the real-world parameters are linear combinations of the risk-neutral
parameters and the market-price-of-risk parameters, we can fit the
risk-neutral and real-world parameters separately and then back out the
market-price-of-risk parameters if we want them. It is just necessary to
keep \(\omega_1=\widetilde{\omega}_1\). This seems to make it easier and
faster for the Stan software. Perhaps Stan uses numerical derivatives of
the posterior probabilities with respect to the parameters as a fitting
step, and having the real-world and risk-neutral parameters related
algebraically complicates this.

The reverting mean of the CIR process is
\(\mu_1 = \omega_1/\kappa_{11}\), as it is for any CIR process. For the
Vasicek processes, the changes to the next period will have mean zero if
all the processes are at their means. A little algebra can give these
means. First, let \(w_2 = \omega_2 - \kappa_{21}\mu_1\) and
\(w_3 = \omega_3 - \kappa_{31}\mu_1\). Also define
\(\kappa_{\Delta} = \kappa_{22}\kappa_{33} - \kappa_{23}\kappa_{32}\).
Then \(\mu_2 = (\kappa_{33}w_2 - \kappa_{23}w_3)/\kappa_{\Delta}\) and
\(\mu_3 = (\kappa_{22}w_3 - \kappa_{32}w_2)/\kappa_{\Delta}\).

The long-term variances of the Vasicek processes, calculated as expected
incremental variance for one year divided by twice the speed of mean
reversion, come out:
\(0.5\sigma_{22}^2\left(\beta_2\omega_1/\kappa_{11}+1\right)/\kappa_{22}\)
and
\(0.5\left[\sigma_{32}^2( \beta_2\omega_1/\kappa_{11}+1) + \sigma_{33}^2(\beta_3\omega_1/\kappa_{11}+1)\right]/\kappa_{33}\).

\textbf{\emph{A1e Unspanned Stochastic Volatility (USV)}}\\
In an \(A_1(3)\) model, or any model that includes a single CIR
component, the current variance of the CIR piece is a constant multiple
of the latest value of that process, and the other variances are linear
functions of the CIR process. That means that the variance of a rate can
be expressed as a linear function of all the rates. This is also the
case when there are multiple CIR processes. The value of the processes
at any point in time can be estimated from the \(C(\tau),D_j(\tau)\)
values at each maturity and the current rates. Thus the variance at any
time can be well estimated by a regression on the rates.

This turns out not to be true of actual rates. Typically a regression
like that has an \(R^2\) of about 20\%. Collin-Dufresne, Goldstein, and
Jones (2003) call this situation ``unspanned stochastic volatility.''
Knowing the yield curve at a given time is not enough to know the
variances of the rates. Since it would be enough in \(A_1(3)\) generated
rates, these models have too close a relationship among the rates and
their volatilities.

Collin-Dufresne, Goldstein, and Jones (2003) then look for affine models
that have stochastic volatility but for which the variance cannot be
computed as a linear function of the rates. They come up with a
closed-form \(A_1(3)\) model with a lot of related parameters which
interact to produce \(D(\tau) = 0\) for the CIR process. Then the rates
are not a linear function of the CIR process, even though its variance
does affect the other processes. Apparently their model did not fit very
well, however. Because only two factors enter into the rate calculation,
it acts more like a two-factor model, for PCA for example.

Joslin (2018) comes up with general constraints for an affine model to
display USV, and gives an \(A_1(4)\) example. Filipovic, Larsson, and
Statti (2018) work out conditions for an \(A_3(3)\) model -- 3 CIRs --
to have USV. USV models do not necessarily fit better -- this is more of
a constraint, like being arbitrage-free. We found in our own fitting
that mistakes in our code that gave models that were not arbitrage-free
often produced better fits. This is typical for constraints. Similarly,
adding more data to the fit usually makes the fit a bit worse for the
original smaller dataset.

\textbf{Appendix 2 -- Fitting models by MCMC}\\
MCMC (Markov Chain Monte Carlo) estimation simulates sample sets of
parameter estimates. It requires a postulated distribution for each
parameter and samples from the conditional distribution of the
parameters given the data, using efficient numerical techniques.
Sometimes it is presented in a Bayesian context, with the postulated
distributions labeled as the priors and the conditional distributions
called posteriors. However these priors often have no connection to any
prior beliefs, or any subjective view of probability, and Bayes Theorem
is not needed to sample from the conditional distributions. The
postulated distributions can be revised in response to the conditional
distributions they generate. MCMC can be done in a frequentist context
if the parameters are instead treated as being random effects with the
postulated distributions. We will use the prior/posterior terminology,
but with the understanding that they are not the same as prescribed by
traditional Bayesian methods, and also have a frequentist
interpretation.

MCMC has goodness of fit measures analogous to the AIC, BIC, etc., the
best one being the leave-one-out (loo) loglikelihood. From the sample of
estimates it is possible to numerically approximate what the likelihood
would be for a point from a fit done with the data excluding that point
-- basically by giving more weight to the parameter sets that fit worst
at this point. Loo is a good estimate of what the loglikelihood would be
for an entirely new sample using the parameters fit to this sample,
which is the goal of the AIC measure as well. All of this is in line
with the idea that model estimation should optimize the fit to the
entire population instead of optimizing the fit to the given sample.

MCMC fitting of yield-curve models includes an intermediate calculation
of the \(r_j(t)\) values for each process \(j\) at each point in time
\(t\). These are not parameters of the model per se, but are still given
priors and produce posteriors. The estimation assumes that there is
noise in the observation process. The interest rates by maturity
produced by a model are the estimated mean values, and the data is
(typically) assumed to be normally distributed with those means and
variance \(\sigma_y\). The likelihood function is calculated from those
probabilities.

In the CVV model, we start with four parameters for each process:
\(\widetilde{\kappa},\widetilde{\omega},\omega\) and \(\sigma\) or
\(\beta\). Those are all given wide-enough priors so that the priors do
not restrict the conditional distributions. Sometimes the priors have to
be narrowed to exclude poor-fitting local maxima, and possibly to speed
the calculations. We use normal priors, but for parameters that must be
positive, we use gamma priors or give their logs normal or uniform
priors. This eliminates a problem with wide priors over-estimating
positive parameters. A similar prior is used for \(\sigma_y\). The prior
for \(\rho\) is initially uniform(\(-1,1\)). These parameters are then
used to calculate the \(C(\tau)\) and \(D_j(\tau)\) functions according
to the formulas above.

The first \(r_j(t)\) for each process is given the prior of the
long-term distribution of the process defined by the parameters, so is
gamma with mean \(\omega/\kappa\) and variance \(\beta \omega/{2}\) for
the CIR, and is normal\((\omega_j/\kappa_j,\sigma_j^2/{2\kappa_j})\) for
the Vasicek processes. This is like assuming that the process has been
going a long time up to that point.

Then the sample value at \(r_j(t)\) is used to produce the prior for the
process at the next period \(r_j(t+dt)\), using the evolution equations
for the processes. Each Vasicek prior for \(r_j(t+dt)\) is normal with
mean \(r_j(t) + [\omega_j - \kappa_j r_j(t)]dt\) and variance
\(\sigma_j^2dt\). The CIR process is approximated by a gamma with mean
and variance
\(\mu = r(t)e^{-\kappa dt} + \kappa\left(1 - e^{-\kappa dt}\right)/\omega\)
and
\(V = \beta^2 \left(1 - e^{-\kappa dt}\right) \left[2r(t)e^{-\kappa dt}+ c \left(1 - e^{-\kappa dt}\right)\right]/2\kappa\).
This has the same mean and variance as the CIR evolution equation, but
for non-instantaneous jumps, the gamma is a better approximation. The
Vasicek priors are bivariate normals with correlation \(=\rho\).

MCMC simulates the conditional distribution of the parameters given the
data. For each simulated set of parameters, it simulates a value of each
\(r_j(t+dt)\) from the parameters and all the \(r_j(t)\)'s for the
processes. Again this gives the conditional parameter distribution given
the data. The different processes can end up with correlated parameters.
Finally the fitted parameters and short rates are taken to be the means
of the conditional distributions.

There are some model diagnostics. Each parameter has a convergence
measure Rhat that will be close to 1.0 if the estimates have converged.
We also check to see if the posterior distributions are pushing against
the boundaries of the priors. If so, the priors are adjusted to
accommodate. There are further diagnostics when parameters do not
converge. Finally, loo can be used to compare alternative models.

\textbf{\emph{A2a Estimation Using Ordinary Differential Equations -
ODEs}}\\
For MCMC model estimation, the prior for \(dr_2(t),r_3(t)\) given
\(r(t)\) is bivariate normal. To get the moments for this, we expand the
evolution matrices. This gives:\\
\[
\begin{aligned}
  \mu_2(t)dt = [\kappa_{21}(\theta_1 - r_1(t))+\kappa_{22}(\theta_2-r_2(t))+\kappa_{23}(\theta_3-r_3(t))]dt \\
 z_2(t)\sqrt{dt} = \sigma_{21} \sqrt{\beta_1r_1(t)dt}\epsilon_1(t)    +\sigma_{22} \sqrt{\alpha_2dt+\beta_2r_1(t)dt}\epsilon_2(t)+\sigma_{23} \sqrt{\alpha_3dt+\beta_3r_1(t)dt}\epsilon_3(t) 
\end{aligned}
\]

and: \[
\begin{aligned}
  \mu_3(t)dt = \large[\kappa_{31}(\theta_1 - r_1(t))+\kappa_{32}(\theta_2-r_2(t))+\kappa_{33}(\theta_3-r_3(t))\large]dt  \\
z_3(t)\sqrt{dt} = \sigma_{31} \sqrt{\beta_1r_1(t)dt}\epsilon_1(t)    +\sigma_{32} \sqrt{\alpha_2dt+\beta_2r_1(t)dt}\epsilon_2(t)+\sigma_{33} \sqrt{\alpha_3dt+\beta_3r_1(t)dt}\epsilon_3(t) 
\end{aligned}
\]

Using \(E[(X-EX)^2]\) for the variance shows that it is the expected
value of the stochastic part squared. All the terms of that that are
mixtures of different \(\epsilon_j\)'s have mean zero. The expected
squared of a mean-zero normal is its variance, so, for a short time
period \(dt\), we have:
\[Variance(dr_2(t)) = \sigma_{21}^2\beta_1r_1(t)dt  +\sigma_{22}^2(\alpha_2+\beta_2r_1(t))dt +\sigma_{23}^2(\alpha_3+\beta_3r_1(t))dt \]
\[Variance(dr_3(t)) = \sigma_{31}^2\beta_1r_1(t)dt  +\sigma_{32}^2(\alpha_2+\beta_2r_1(t))dt +\sigma_{33}^2(\alpha_3+\beta_3r_1(t))dt \]

These depend on \(r_(t)\), the CIR process, but for an \(A_1(3)\) model,
the variances do not depend on the Vasicek processes.

We calculate the covariance using \(E[(X-EX)(Y - EY)]\). This is the
product of the two stochastic terms, and again any mixed products have
mean zero. This gives the incremental covariance:
\[Cov(dr_2(t),dr_3(t)) = \sigma_{21} \sigma_{31} \beta_1r_1(t)dt+ \sigma_{22} \sigma_{32} (\alpha_2 + \beta_2r_1(t))dt + \sigma_{23} \sigma_{33} (\alpha_3 + \beta_3r_1(t))dt\]

For the prior for the starting point of each process, we again assume
that each one has the long-term distribution for the process. The mean
for each is its \(\theta_j\), the reverting mean. The variances are the
one-year variances (i.e., for \(dt=1\)) divided by twice the speed of
mean reversion \(\kappa_{jj}\), where for this purpose, \(r_1\) takes
its mean value \(\theta_1\). By convention, mean reversion is expressed
in annual terms.

Stan has a solver for systems of differential equations. You write the
system as a single function at the top of the code, in the functions
section. It gives as output the vector of left-hand sides of the system
-- here the \(d/d\tau\) terms. We need \(C(\tau)\) and \(D_j(\tau)\)
functions for each maturity \(\tau\), but for a given set of parameters,
these are fixed across the observation times. Thus the function takes as
arguments the current values of:
\(\tau, \widetilde{K}, \widetilde{\Omega}, \Sigma, \beta0_j, j=1,2,3\).
This does not require \(r(t)\). Then to solve the system, the
differential-equation solver function is applied to the output of the
differential equation function. For some reason, Stan's name for this
solver is ``integrate\_ode\_rk45.''

\textbf{Appendix 3 -- Code}\\
Stan and R code is up on the CAS GitHub site for the E-Forum, at
\url{https://github.com/casact/ef_yield_curve_generators}. All the files
can be downloaded by clicking the ``Clone or download'' button at the
middle right of the webpage. If you have not joined GitHub, this will
still work but you will be prompted to join. The code should all go into
one folder and that should be declared as the Working Directory for R.
There are GitHub disclaimers at the top of each file, but these should
not affect the implementation. The files are:

\begin{itemize}
   \item CIR.R -- R code to start the run.
    \item us\_01\_05\_18\_to\_10\_04\_19.csv -- csv file that includes historical yields from January 5, 2018 to October 4, 2019. This might be updated occasionally, and the users can update it by adding more rows.
      \item CIR.stan -- Stan file for CIR model with closed form solutions.
  \item CIR\_ode.stan -- Stan file for CIR model with ordinary differential equations. Only there to show the ode code for a simple model.
  \item CVV\_ plus.stan -- Stan file for CVV+ model.
  \item 7k3b.stan -- Stan file for 7k3b model.
\end{itemize}

By default, the R code will use CIR\_ode.stan as the model. Other models
can be used by changing the stan file name in line 24. The code is just
something that works and is undoubtedly not optimal.

The introductory Stan file is that for a single CIR process. The
structure of a Stan file is to first read in variables already populated
from the R space, then define all the variables to be used. That takes a
fair amount of real estate in a Stan file, and is needed because it will
be translated to C++ then compiled. We make the risk-neutral parameters
the first ones to estimate, and the model begins with calculating the
\(C\) and \(D\) functions. Then the real-world parameters are defined,
and from these the processes at each point are estimated. In the CIR
file that just uses the CIR evolution equation to define the prior for
the CIR process at each time point, conditional on the previous point.
The gamma distribution approximation is used for that. From all this,
the fitted values are calculated, then the model for the data is just
normal in these values. Finally the likelihoods are computed to pass to
loo.

We also put up this model done by solving the differential equations, as
an example of how to do that, although no one ever would as the
equations have already been solved in closed form. To do it, you start
with a new functions section above the data section, and define the
system of differential equations there. We call that system AB\_eq, as
it solves for the A and B functions. These produce the C and D functions
after the system has been solved. It is solved by the function
integrate\_ode\_rk45. This takes as arguments the name of the system to
be solved, starting values at \(\tau=0\) -- here a vector of zeros, the
values of the risk-neutral parameters, the precision wanted for the
iteration, and some other arguments. The latter are not well documented
and are required even though not used, so we put in some (obscure)
values that we found in Stan examples, and it all seems to work well.
The values of the parameters are strung out in a linear array to pass to
the function, which then puts them back into vectors, etc. There is
probably a simpler way to do that, but we got it working by doing it, so
kept doing it that way. The intermediate W and h calculations for CIR
are not needed here, as those are for solving the system in closed form.
All the other parts of the code after getting the C and D functions are
the same as in the closed-form case.

For the three-factor models, the Vasicek formulas and the correlation
adjustment are needed. There is code provided for the CVV+ essentially
affine model with the \(\delta\) and \(\gamma\) terms included, and for
the 7k3b model. The correlation adjustment is not needed for the 7k3b
model, since it solves for C and D numerically. The priors for it are
what gave the fit above, but alternative priors are shown as comments.
They gave an even better fit, but the priors are very narrow, and with
much deviation from these ranges the model deteriorates rapidly. That
makes them suspect as a fluke set of priors that works for the current
data but is not really representative of a longer-term population.
Still, both sets are worth trying. Probably for new data they would both
have to be modified after seeing how they perform. The posterior
histograms can help show if the parameters are trying to move in one
direction or another from the priors used.

Some miscellaneous R code is shown below for defining the paras variable
used in the print and plot statements, and for extracting the simulated
samples for use in simulating future scenarios.

\begin{Shaded}
\begin{Highlighting}[]
\CommentTok{#for 7k3b model}
\NormalTok{paras <-}\StringTok{ }\KeywordTok{c}\NormalTok{(}\StringTok{"kaprn"}\NormalTok{,}\StringTok{"kaprn21"}\NormalTok{,}\StringTok{"kaprn23"}\NormalTok{,}\StringTok{"kaprn31"}\NormalTok{,}\StringTok{"kaprn32"}\NormalTok{,}\StringTok{"omrn"}\NormalTok{,}\StringTok{"lb1"}\NormalTok{,}\StringTok{"lb2"}\NormalTok{,}
\StringTok{"lb3"}\NormalTok{,}\StringTok{"ls2"}\NormalTok{,}\StringTok{"ls3"}\NormalTok{, }\StringTok{"corr"}\NormalTok{,}\StringTok{"kap"}\NormalTok{, }\StringTok{"kap21"}\NormalTok{,}\StringTok{"kap23"}\NormalTok{,}\StringTok{"kap31"}\NormalTok{,}\StringTok{"kap32"}\NormalTok{,}\StringTok{"om"}\NormalTok{,}\StringTok{"ldel"}\NormalTok{,}\StringTok{"gam1"}\NormalTok{) }

\CommentTok{#for CVV_plus model}
\NormalTok{paras <-}\StringTok{ }
\KeywordTok{c}\NormalTok{(}\StringTok{"kaprn"}\NormalTok{, }\StringTok{"omrn"}\NormalTok{, }\StringTok{"lb"}\NormalTok{, }\StringTok{"ls2"}\NormalTok{, }\StringTok{"ls3"}\NormalTok{, }\StringTok{"sigma_y"}\NormalTok{, }\StringTok{"corr"}\NormalTok{, }\StringTok{"kap"}\NormalTok{, }\StringTok{"om"}\NormalTok{,}\StringTok{"ldel"}\NormalTok{,}\StringTok{"gam1"}\NormalTok{)}

\CommentTok{#extracting scenarios}
\NormalTok{us1_ss =}\StringTok{ }\KeywordTok{extract}\NormalTok{(us_}\DecValTok{1}\NormalTok{, }\DataTypeTok{permuted =} \OtherTok{FALSE}\NormalTok{) }\CommentTok{# this gets all the samples}
\CommentTok{#Need permuted = FALSE to get it in array form}
\KeywordTok{dim}\NormalTok{(us1_ss) }\CommentTok{# shows dimensions, like 1000 x 4 x 2000 for 1000 sampling draws, }
\CommentTok{#4 chains and 2000 things computed}
\NormalTok{us1_ss =}\StringTok{ }\NormalTok{us1_ss[,,}\KeywordTok{c}\NormalTok{(}\DecValTok{1}\OperatorTok{:}\DecValTok{28}\NormalTok{,}\DecValTok{105}\NormalTok{,}\DecValTok{182}\NormalTok{,}\DecValTok{259}\OperatorTok{:}\DecValTok{277}\NormalTok{,}\DecValTok{286}\OperatorTok{:}\DecValTok{336}\NormalTok{)] }
\CommentTok{# keeps only variables that are needed for 7k3b;}
\CommentTok{#for simulation. Here 105, 182, 259 were the last (90th) period's r values}
\CommentTok{#these numbers would change with more periods fit; also taus (278-285) left out}
\CommentTok{#careful though as arrays show in different order than in print output}
\CommentTok{#just keeping 1 chain will keep variables names to check: us1_ss = us_1_ss[,1,]}
\KeywordTok{dim}\NormalTok{(us1_ss) =}\StringTok{ }\KeywordTok{c}\NormalTok{(}\DecValTok{4000}\NormalTok{,}\DecValTok{100}\NormalTok{) }\CommentTok{# collapses first two dimensions (samples and chains)}
\KeywordTok{write.csv}\NormalTok{(us1_ss, }\DataTypeTok{file=}\StringTok{"samples_out.csv"}\NormalTok{)}
\end{Highlighting}
\end{Shaded}

\textbf{References}

\hypertarget{refs}{}
\leavevmode\hypertarget{ref-bolder2001}{}%
Bolder, David Jamieson. 2001. ``Affine Term-Structure Models: Theory and
Implementation.'' \emph{Bank of Canada Working Paper}
https://www.bankofcanada.ca/wp-content/uploads/2010/02/wp01-15a.pdf:
2001--15.

\leavevmode\hypertarget{ref-brigo2006}{}%
Brigo, Damiano, and Fabio Mercurio. 2001. ``Interest Rate Models -
Theory and Practice.'' \emph{Springer}.

\leavevmode\hypertarget{ref-campbell1991}{}%
Campbell, J. Y., and R. J. Shiller. 1991. ``Yield Spreads and Interest
Rate Movements: A Bird's Eye View.'' \emph{Review of Economic Studies}
58: 495--514.

\leavevmode\hypertarget{ref-christensen2015}{}%
Christensen, Jens H. E. 2015. ``Affine Term Structure Models: An
Introduction.'' \emph{European University Institute}.

\leavevmode\hypertarget{ref-collin2003}{}%
Collin-Dufresne, Pierre, Robert S. Goldstein, and Christopher S. Jones.
2003. ``Identification and Estimation of `Maximal' Affine Term Structure
Models: An Application to Stochastic Volatility''
https://pdfs.semanticscholar.org/5184/b8b8d761027012c42a28f84abfbb58e345db.pdf.

\leavevmode\hypertarget{ref-cox1985}{}%
Cox, John C., Jonathan E. Ingersoll, and Stephen A. Ross. 1985. ``A
Theory of the Term Structure of Interest Rates.'' \emph{Econometrica}
53: 385--408.

\leavevmode\hypertarget{ref-dai2000}{}%
Dai, Qiang, and Kenneth J. Singleton. 2000. ``Specification Analysis of
Affine Term Structure Models.'' \emph{Journal of Finance} 55:5:
1943--78.

\leavevmode\hypertarget{ref-feldhutter2016}{}%
Feldhütter, Peter. 2016. ``Can Affine Models Match the Moments in Bond
Yields?'' \emph{Quarterly Journal of Finance} 6:2:
http://www.feldhutter.com/RiskPremiumPaperFinal.pdf.

\leavevmode\hypertarget{ref-filip2018}{}%
Filipovic, Damir, Martin Larsson, and Francesco Statti. 2018.
``Unspanned Stochastic Volatility in the Multifactor Cir Model.''
\emph{Mathematical Finance} https://arxiv.org/pdf/1705.02789.pdf:
26/September/2018.

\leavevmode\hypertarget{ref-jag2003}{}%
Jagannathan, R., A. Kaplin, and S. Sun. 2003. ``An Evaluation of
Multi-Factor Cir Models Using Libor, Swap Rates and Cap and Swaption
Prices.'' \emph{Journal of Econometrics} 116: 113--46.

\leavevmode\hypertarget{ref-joslin2018}{}%
Joslin, Scott. 2018. ``Can Unspanned Stochastic Volatility Models
Explain the Cross Section of Bond Volatilities?'' \emph{Management
Science}, 64:4.

\leavevmode\hypertarget{ref-pedersen2016}{}%
Pedersen, Hal, Mary Pat Campbell, Stephan L. Christiansen, Samuel H.
Cox, Daniel Finn, Ken Griffin, Nigel Hooker, Matthew Lightwood, Stephen
M. Sonlin, and Chris Suchar. 2016. ``Economic Scenario Generators a
Practical Guide.'' \emph{Society of Actuaries}.

\leavevmode\hypertarget{ref-piketty2014}{}%
Piketty, Thomas. 2014. ``Capital in the Twenty-First Century.''
\emph{Belknap Press} Cambridge, MA.

\leavevmode\hypertarget{ref-troiani2017}{}%
Troiani, Angelo. 2017. ``Interest Rate Models: Calibration and
Validation Applying Kalman Filter.'' \emph{Italian Actuaries
Organization}
https://www.italian-actuaries.org/wp-content/uploads/2017/10/Vasicek2-Calibration-and-Validation.pdf.

\leavevmode\hypertarget{ref-vasicek1977}{}%
Vasicek, Oldrich A. 1977. ``An Equilibrium Characterization of the Term
Structure.'' \emph{Journal of Financial Economics} 5: 177--88.

\leavevmode\hypertarget{ref-venter2004}{}%
Venter, Gary. 2004. ``Testing Distributions of Stochastically Generated
Yield Curves.'' \emph{Astin Bulletin} 34:1:
https://www.actuaries.org/LIBRARY/ASTIN/vol34no1/229.pdf.

\leavevmode\hypertarget{ref-ye1998}{}%
Ye, J. 1998. ``On Measuring and Correcting the Effects of Data Mining
and Model Selection.'' \emph{Journal of the American Statistical
Association} 93: 120--31.

\end{document}